\documentclass[12pt]{article}

\usepackage[a4paper,left=30mm,right=20mm,top=20mm,bottom=20mm]{geometry}
\usepackage{graphics}
\usepackage{amsmath}
\usepackage{epsfig}
\usepackage{cite}

\title{Employing RHIC and LHC data to determine TMD gluon density in a proton}

\author{N.A.~Abdulov$^{1}$, H.~Jung$^2$, A.V.~Lipatov$^{3,\,4}$, G.I.~Lykasov$^4$, M.A.~Malyshev$^3$}

\begin{document}

\maketitle

\begin{center}

{\it $^1$Faculty of Physics, Lomonosov Moscow State University, 119991 Moscow, Russia}\\
{\it $^2$Deutsches Elektronen-Synchrotron, 22603 Hamburg, Germany}\\
{\it $^3$Skobeltsyn Institute of Nuclear Physics, Lomonosov Moscow State University, 119991 Moscow, Russia}\\
{\it $^4$Joint Institute for Nuclear Research, 141980 Dubna, Moscow Region, Russia}

\end{center}

\vspace{0.5cm}

\begin{center}

{\bf Abstract }

\end{center} 

\indent 
Transverse momentum dependent (TMD) parton distributions in a proton 
are important in high energy physics from both theoretical and phenomenological
points of view. 
Using the latest RHIC and LHC data on the inclusive soft hadron production 
in $pp$ and $AA$ collisions at small
transverse momenta, we determine the parameters of the initial TMD gluon density,
derived in the framework of quark-gluon string model
at the low scale $\mu_0 \sim 1 - 2$~GeV and
refine its large-$x$ behaviour
using the LHC data on the $t \bar t$ production at $\sqrt s = 13$~TeV.
Then, we 
apply the
Catani-Ciafaloni-Fiorani-Marchesini (CCFM) evolution 
equation to extend the obtained TMD gluon density to the whole kinematical region.
In addition, the complementary TMD valence and sea quark distributions
are generated. The latter are evaluated
in the approximation where the gluon-to-quark splitting occurs at 
the last evolution step
using the TMD gluon-to-quark splitting function.
Several phenomenological applications of the proposed
TMD quark and gluon densities to the LHC processes are discussed.

\vspace{1.0cm}

\noindent
PACS number(s): 12.38.Bx, 14.65.Dw, 14.65.Fy

\newpage

\section{Introduction} \indent

In the recent years an understanding has been obtained that a theoretical
description of a number of processes at high energies and large momentum transfer
containing multiple hard scales requires unintegrated, or transverse momentum
dependent (TMD) parton density functions\cite{1}, which
encode non-perturbative information on proton structure, including transverse momentum
and polarization degrees of freedom.
The TMD parton densities are related to the physical
cross sections and other observables, measured in the collider experiments,
via TMD factorization theorems in Quantum Chromodynamics (QCD).
At present, the Collins-Soper-Sterman factorization approach, designed for 
semi-inclusive processes with a finite and non-zero ratio
between the hard scale and total energy\cite{2}, 
and high-energy factorization\cite{3} (or $k_T$-factorization\cite{4}) approach,
valid in the limit of a fixed hard scale and high energy,
are developed.
The factorization theorems provide the necessary framework to
separate hard partonic physics, described with a perturbative QCD
expansion, from soft hadronic physics and allow one to determine 
the TMD parton distributions from collider data.

In the high-energy factorization, the production cross sections
at low transverse momenta are governed by the non-perturbative input to 
the TMD parton density functions. The latter, being used as an initial condition
for the subsequent QCD evolution, could play an important role for phenonemological applications\cite{5,6,7,8}.
As it was shown\cite{9,10,11}, its influence on the description of the experimental data can be
significant.
At present, several approaches to determine
the TMD gluon density in a proton (or rather its initial parameters)
are known in the literature.
In the Kimber-Martin-Ryskin (KMR) scheme, developed at 
leading (LO)\cite{12} and next-to-leading (NLO)\cite{13} orders,
the TMD quark and gluon densities are derived from the conventional
parton distribution functions.
At low $k_T \leq \mu_0 \sim 1$~GeV they are defined
from a simple normalization condition.
Recently, the TMD quark and gluon densities in a proton were determined\cite{14} from fits to precision measurements of deep 
inelastic scattering cross sections at HERA and
evolved by Dokshitzer-Gribov-Lipatov-Altarelli-Parisi (DGLAP) evolution\cite{15} with NLO 
splitting functions using the parton branching method\cite{16,17}.
In a more complicated approach\cite{18},
based on the Catani-Ciafaloni-Fiorani-Marchesini (CCFM) gluon evolution 
equation\cite{19},
the parameters of initial TMD gluon distribution
were fitted from the precision HERA data on proton structure 
function $F_2(x,Q^2)$ in the range $x < 5\cdot 10^{-3}$, $Q^2 > 5$~GeV$^2$ 
and $F_2^c(x,Q^2)$ at $Q^2 > 2.5$~GeV$^2$, assuming the Gaussian-like dependence on the intrinsic gluon
transverse momentum $k_T$ at $k_T \leq \mu_0 \sim 2$~GeV.
In our previous papers\cite{9,10,11}
the initial TMD gluon density was derived in the framework of the soft quark-gluon string model\cite{20,21,22}
by taking into account gluon saturation effects at low $Q^2$.
The essential parameters were obtained from the best description 
of the inclusive spectra of hadrons produced in $pp$ collsions 
at LHC energies in the midrapidity region at low transverse momenta $p_T \leq 4.5$~GeV.
Being used with the CCFM evolution, 
the predictions based on the proposed TMD gluon density describe well the HERA 
measurements of the proton structure functions $F_2^c(x,Q^2)$, $F_2^b(x,Q^2)$ and $F_L(x,Q^2)$.
Thereby, the connection between soft LHC processes and small-$x$ physics at HERA
in a wide kinematical region was established.
An important advantage of the approaches\cite{10,11,18} is that 
one can rather easily take into account a large piece of 
higher-order corrections, namely, part of NLO + NNLO + ...  terms containing 
leading $\log 1/x$ enhancement of cross sections due to real
initial state parton emissions, absorbed into the CCFM 
evolution\footnote{At present, most of the proposed TMD parton distributions in a proton is 
collected in the \textsc{tmdlib} package\cite{23}, which is a C++ library 
providing a framework and an interface to the different parametrizations.} (see\cite{24} for more information).

In the present paper we continue our previous studies\cite{9,10,11} and 
test the parameters of the initial TMD gluon density\cite{9,10,11}
using the recent NA61\cite{25}, LHC\cite{26,27,28,29,30,31} and RHIC\cite{32,33} data 
for soft hadron production in $pp$ and $AA$ collisions obtained in a wide 
energy range.
Moreover, we refine its large-$x$ behaviour
using the latest LHC data on the inclusive top quark pair production at $\sqrt s = 13$~TeV\cite{34}.
Following Refs.\cite{10,11}, we extend the updated TMD gluon distribution to the whole range 
of the longitudinal momentum fraction $x$, transverse momentum ${\mathbf k_T^2}$ and hard scale $\mu^2$
numerically 
using the \textsc{updfevolv} package\cite{35}, which is 
the CCFM evolution code for TMD parton densities.
In our opinion, the CCFM equation is the most suitable tool
for our study since it smoothly interpolates between the 
small-$x$ Balitsky-Fadin-Kuraev-Lipatov\cite{36} (BFKL) gluon dynamics
and the conventional DGLAP one.
We supply the obtained TMD gluon density with the corresponding
TMD valence and sea quark distributions, calculated in the approximation, where 
the sea quarks occur in the last gluon splitting.
Finally, we discuss several phenomenological applications
of the proposed TMD parton densities to hard LHC processes, sensitive to
the quark and gluon content of the proton.

The paper is organized as follows. In Section~2 we describe how we determine the 
initial gluon density from the LHC data and discuss its subsequent QCD evolution.
In Section~3 we illustrate the use of the obtained TMD gluon density at the LHC.
We give conclusions in Section~4.

\section{Non-perturbative TMD gluon input and evolution} \indent

In fact, the determination of the parameters of the initial TMD gluon density in a proton
can be splitted into the two almost independent pieces refering to 
the regions of small and large $x$. We consider first the small-$x$ region and
start from the simple analytical expression for 
the starting TMD gluon distribution function at some fixed scale $\mu_0 \sim 1 - 2$~GeV. It can be presented
in a form\cite{11}
\begin{equation}
  f_g^{(0)}(x,{\mathbf k}_T^2,\mu_0^2) = \tilde f_g^{(0)}(x,{\mathbf k}_T^2,\mu_0^2) + \lambda(x, {\mathbf k}_T^2,\mu_0^2) f_g(x,{\mathbf k}_T^2),
\end{equation}

\noindent 
where $x$ and ${\mathbf k}_T$ are the proton longitudinal momentum fraction and two-dimensional gluon 
transverse momentum, respectively. The first term, $\tilde f_g^{(0)}(x,{\mathbf k}_T^2,\mu_0^2)$, was 
calculated\cite{9} within the soft QCD model and reads:
\begin{equation}
  \displaystyle \tilde f_g^{(0)}(x,{\mathbf k}_T^2,\mu_0^2) = c_0 c_1 (1-x)^{b} \times \atop {
  \displaystyle \times \left[R_0^2(x){\mathbf k}_T^2 + c_2\left(R_0^2(x){\mathbf k}_T^2\right)^{a/2}\right] \exp\left(-R_0(x)
  |{\mathbf k}_T|-d\left[R_0^2(x){\mathbf k}_T^2\right]^{3/2}\right)},
\end{equation}

\noindent
where $R_0^2(x) = (x/x_0)^\lambda/\mu_0^2$ and $c_0 = 3 \sigma_0/4\pi^2 \alpha_s$. 
The parameters $\sigma_0 = 29.12$~mb, $\lambda = 0.22$, $x_0 = 4.21 \cdot 10^{-5}$ and $\alpha_s = 0.2$
come from the Golec-Biernat-W\"usthoff (GBW) saturation model\cite{37}, while other parameters
$a$, $b$, $c_1$, $c_2$ and $d$
were fitted from LHC data on inclusive spectra of charged hadrons.
The numerical values of these parameters, details of the calculations and the relation 
between the TMD gluon density and the inclusive hadron spectra are given in
our previous papers\cite{9,10,11}.
The gluon density $\tilde f_g^{(0)}(x,{\mathbf k}_T^2,\mu_0^2)$ differs from the one obtained 
in the GBW model at $|{\mathbf k}_T| < 1$~GeV and coincides with the GBW gluon 
at larger $|{\mathbf k}_T| > 1.5$~GeV\cite{9}. The second term, $f_g(x,{\mathbf k}_T^2)$, represents 
the analytical solution\cite{38} of the linear BFKL equation at low $x$ weighted with a
matching function $\lambda(x, {\mathbf k}_T^2,\mu_0^2)$:
\begin{equation}
  f_g(x,{\mathbf k}_T^2) = \alpha_s^2 \, x^{-\Delta} \, t^{-1/2} {1 \over v} \exp \left[ - { \pi \ln^2 v \over t}\right],
\end{equation}
\begin{equation}
  \lambda(x,{\mathbf k}_T^2,\mu_0^2) = c_0 \left({x\over x_0}\right)^{0.81} \exp\left[ - k_0^2 {R_0(x)\over |{\mathbf k}_T|} \right],
\end{equation}

\noindent 
where $t = 14\,\alpha_s N_c \, \zeta(3) \ln(1/x)$, $\Delta = 4 \, \alpha_s N_c \ln 2 /\pi$, 
$v = |{\mathbf k}_T|/\Lambda_{\rm QCD}$ and $k_0 = 1$~GeV. This term allows one 
to describe LHC measurements of inclusive charged hadrons
up to $p_T \leq 4.5$~GeV\cite{11}. It is important that the contribution from 
$f_g(x,{\mathbf k}_T^2)$ is only non-zero at $|{\mathbf k}_T| \ll \Lambda_{\rm QCD} \, (1/x)^{\delta}$
with $\delta = \alpha_s N_c$, resulting in an average generated gluon transverse momentum of
$\langle |{\mathbf k}_T| \rangle \sim 1.9$~GeV. The latter value is close 
to the non-perturbative QCD regime, that allows one to treat the TMD gluon density above
as a starting one for the CCFM evolution.

Previously, the phenomenological parameters $a$, $b$, $c_1$, $c_2$ and $d$ in (1) --- (4) were determined
in the small-$x$ region only, where $x \sim 1 \cdot 10^{-4}$ --- $1 \cdot 10^{-5}$ (see\cite{9,10,11}).
The fit was based on NA61 data on inclusive cross sections of $\pi^-$ meson production in $pp$ 
collisions at initial momenta $31$ and $158$ GeV\cite{25} and on CMS\cite{26} and ATLAS\cite{27} data on inclusive hadron production
in $pp$ collisions at the LHC.
In the present note we 
tested all these parameters using
the experimental data on the pion tranverse mass distribution in Au + Au and Pb + Pb collisions 
taken by the STAR Collaboration at the RHIC\cite{32,33}
and ALICE Collaboration at the LHC\cite{28,29,30,31}.
The details of the calculations of hadron production cross sections 
in $AA$ collisions are given in\cite{39}. 
Let us stress that 
the possible higher-order corrections (see, for example,\cite{40,41,42})
to the leading-order BFKL motivated $k_T$-dependence of the proposed gluon 
input at low-$x$ (as well as saturation dynamics) are effectively included.

Next, we note that determination of the parameters of the TMD gluon density in the
small-$x$ region only could result in significant theoretical uncertainties of the predictions and/or poor 
description of the data at moderate and large $x$ values.
Therefore, in the present paper we refine some of these parameters,
essential in the large $x$ region,
using recent experimental data on inclusive $t\bar t$ production taken by the CMS
Collaboration at $\sqrt s = 13$~TeV\cite{34}. These data refer 
to $x \sim 2 m_t/\sqrt s \sim 3 \cdot 10^{-2}$ (with a top mass $m_t \sim 170$~GeV)
and are reported at the parton level in the full phase space, 
allowing us to avoid the numerical simulation of top quark decays.
To calculate the $t \bar t$ production cross sections in
the $k_T$-factorization approach we follow our previous consideration\cite{43}.
We find that $b = 10$ and $d = 0.4$ are 
more preferable to describe the distributions on the rapidity and 
transverse momentum of top quark pairs.
The latter leads, in addition, to a different value of overall normalization $n = 0.27$ in~(1),
which was determined using the CMS data on inclusive $b$-jet production.

\begin{figure}
\begin{center}
\includegraphics[width=7.9cm]{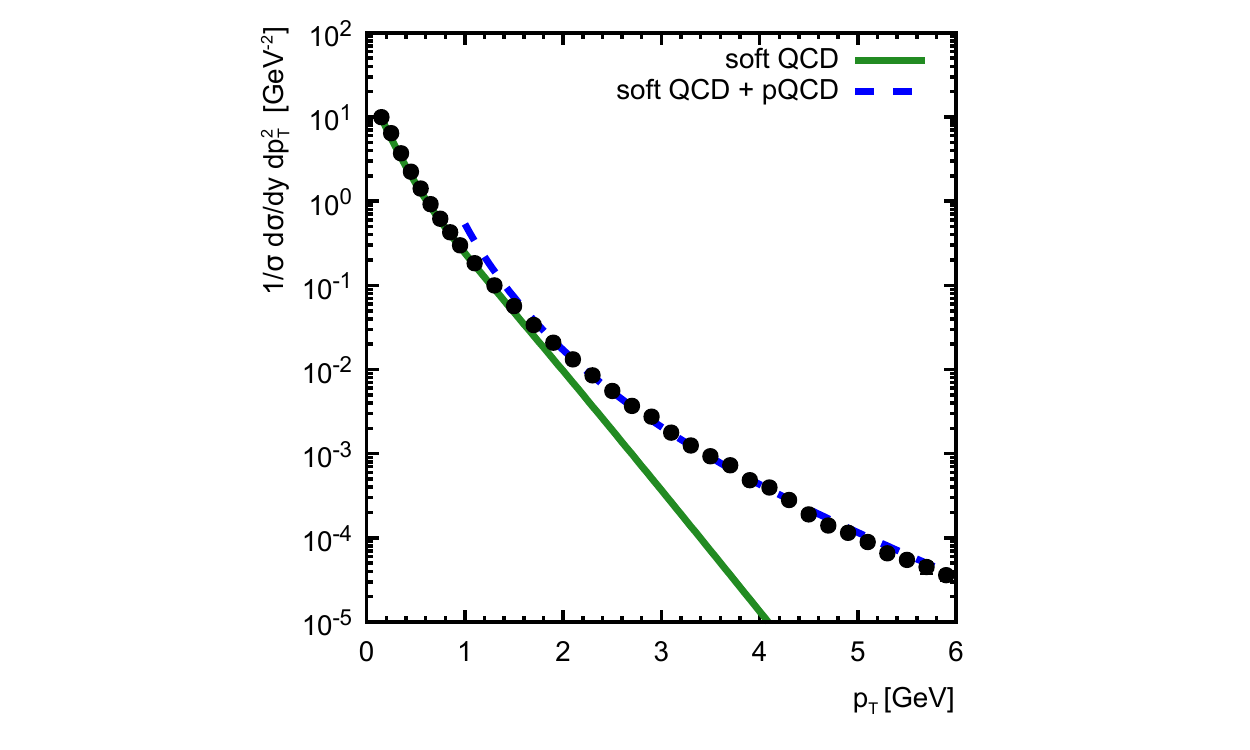}
\includegraphics[width=7.9cm]{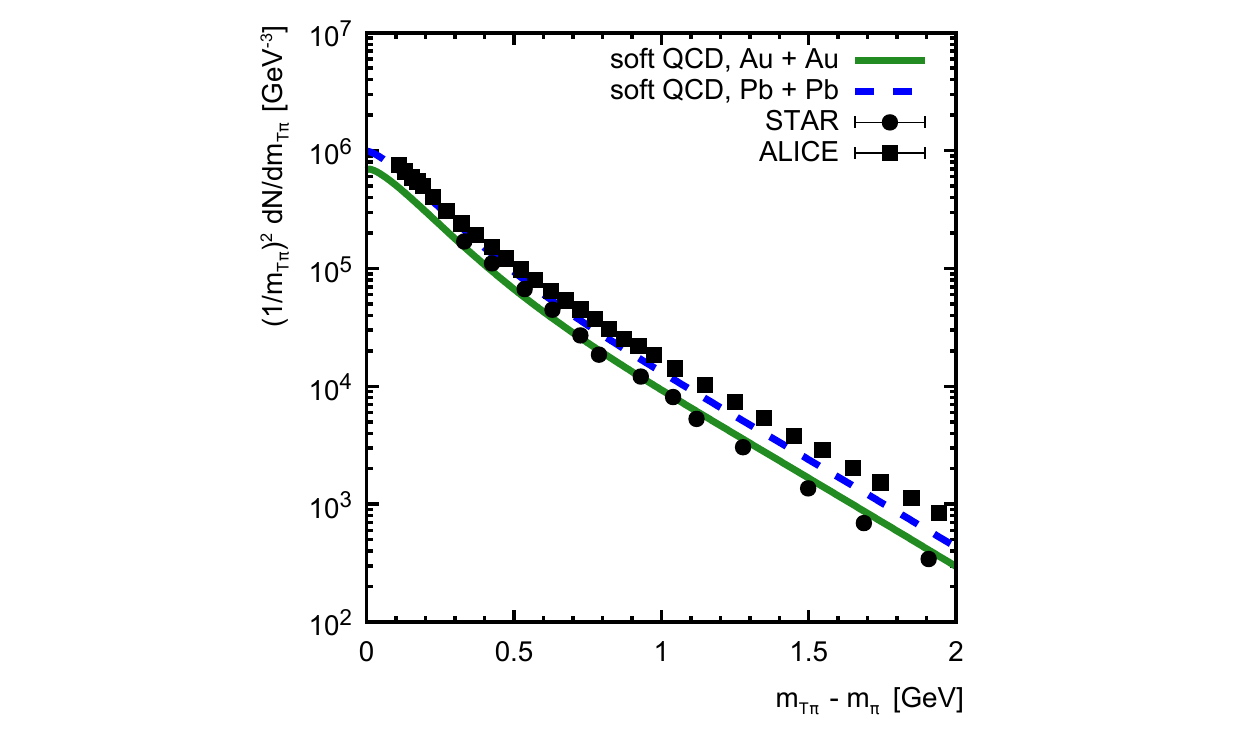}
\caption{Left panel: the inclusive cross section of charge hadrons
produced in $pp$ collisions as a function of their transverse momentum
at $\sqrt{s}=$7~TeV. The experimental data are from CMS and ATLAS\cite{26,27}.
Right panel: pion transverse mass spectra in Au + Au and
Pb + Pb collisions.
The experimental data are from STAR\cite{32,33} and ALICE\cite{28,29,30,31}.}
\label{fig1}
\end{center}
\end{figure}

The illustration of the satisfactory description of the RHIC\cite{32,33} and LHC data\cite{28,29,30,31} 
on soft hadron production in
$pp$ and $AA$ collisions at mid-rapidities is presented in Fig.~\ref{fig1}.
The soft QCD predictions include both gluon and quark contributions.
The perturbative QCD corrections, calculated\cite{44,45} at LO, are divergent at low transverse
momenta\footnote{The kinematical region $p_T \sim 1.8 - 2.2$~GeV can be treated as 
the matching region of the soft QCD and pQCD calculations.}
(not shown for $AA$ collisions).
The hadron production at $p_T < 2$~GeV are fitted with $\chi^2/n.d.f = 0.9$.
We would like to note here that the approach\cite{39} with the above determined 
parameters of the TMD gluon density is able to describe the 
experimental data in a wide energy range.
Concerning the large-$x$ region, the achieved description of the CMS data\cite{34} on the top pair production
is illustrated in Fig.~2,
where the transverse momentum and rapidity distributions of the top quarks are shown as an example.
For the reader's convenience, we collected all the parameters of (1) --- (4) 
in Table~1.

\begin{table}
\begin{center}
\begin{tabular}{|l|c|c|c|c|c|c|}
\hline
   & & & & & & \\
   Parameter & $a$ & $b$ & $c_1$ & $c_2$ & $d$ & $\mu_0/$GeV\\
   & & & & & & \\
\hline
   & & & & & & \\
   Fitted value & $0.3$ & $10.0$ & $0.3295$ & $2.3$ & $0.4$ & $2.2$\\
   & & & & & &\\
\hline
\end{tabular}
\end{center}
\caption{Numerical values of the parameters of the TMD gluon density (1) --- (4). All other 
parameters, namely, $x_0$, $\sigma_0$, $\lambda$ and $\alpha_s$ are unchanged.}
\label{table1}
\end{table}

Next, we extend the obtained TMD gluon density~(1) --- (4) to a higher scale $\mu^2$ using
the CCFM evolution equation. This equation resums 
large logarithms $\alpha_s^n \ln^n 1/x$ and $\alpha_s^n \ln^n 1/(1-x)$ 
and, therefore, is valid at both small and large $x$ (see, for example,\cite{24} for more information).
In the leading logarthmic approximation\footnote{The next-to-leading logarthmic corrections for the 
CCFM equation are still unknown. However, as it was argued\cite{46}, amending the 
leading logarithmic evolution with kinematical constraint\cite{47,48} leads to reasonable
QCD predictions, although still formally only to leading logarithmic accuracy (see also\cite{24}).}, the CCFM equation with 
respect to the evolution scale $\mu^2$ can be written as
\begin{equation}
\displaystyle f_g(x,{\mathbf k}_T^2,\mu^2) = f_g^{(0)}(x,{\mathbf k}_T^2,\mu_0^2) \Delta_s(\mu^2,\mu_0^2) + \atop { 
\displaystyle + \int {dz\over z} \int {d q^2\over q^2} \theta(\mu - z q) \Delta_s(\mu^2,z^2 q^2)} P_{gg}(z,q^2,{\mathbf k}_T^2) 
f_g(x/z,{\mathbf k^\prime}_T^2,q^2),
\end{equation}

\noindent 
where ${\mathbf k^\prime}_T = {\mathbf q} (1 - z) + {\mathbf k}_T$. The
exact analytical expressions for the Sudakov form factor 
$\Delta_s(p^2, q^2)$ and 
gluon splitting functions $P_{gg}(z,q^2,{\mathbf k}_T^2)$
can be found, for example, in\cite{35}.
The CCFM equation with the starting TMD gluon
density $f_g^{(0)}(x,{\mathbf k}_T^2,\mu_0^2)$ given by (1) --- (4) was solved numerically
using the \textsc{updfevolv} package\cite{35}.
As it was done earlier\cite{11}, to produce the TMD valence and sea quark distributions
we apply the approach\cite{49}. So, the TMD sea quark density was calculated 
in the approximation where the sea quarks 
occur in the last gluon splitting:
\begin{equation}
  f_{q}^{(s)}(x,{\mathbf k}_T^2,\mu^2) = \int \limits_x^1 {dz \over z} \int d{\mathbf q}_T^2
    {1\over {\mathbf \Delta}^2} {\alpha_s \over 2\pi} P_{qg}(z,{\mathbf q}_T^2,{\mathbf \Delta}^2) f_g(x/z,{\mathbf q}_T^2, \bar \mu^2),
\end{equation}

\noindent
where $z$ is the fraction of the gluon light-cone momentum carried by
the quark and $\mathbf \Delta = {\mathbf k}_T - z{\mathbf q}_T$. 
The hard scale $\bar \mu^2$ was defined\cite{50} from the angular ordering condition which is natural
from the CCFM point of view: $\bar \mu^2 = {\mathbf \Delta}^2/(1-z)^2 + {\mathbf q}_T^2/(1-z)$.
The off-shell gluon-to-quark splitting function $P_{qg}(z,{\mathbf q}_T^2,{\mathbf \Delta}^2)$
was calculated in\cite{51}.

The gluon density $f_g(x,{\mathbf k}_T^2,\mu^2)$
obtained according to~(1) --- (5),
labeled below as {\it Moscow-Dubna 2018}, or {\it MD'2018},
is shown in Fig.~3 versus the longitudinal momentum fraction $x$ and
transverse momentum ${\mathbf k}_T$ at different evolution scales.
Additionally, we plot the TMD gluon distribution\cite{18} (namely, the {\it JH'2013 set 2})
which is widely discussed in the literature and commonly used in applications. 
One can observe some difference in the absolute normalization and shape between 
both TMD gluon distributions. In particular, the ${\mathbf k}_T$-tail of the {\it MD'2018} density function
is the contribution due to the solution of the linear BFKL equation, as was described above. 
Actually, it was needed to improve the description of the LHC
data on charge hadron production in $pp$ collisions at $\sqrt{s} = 7$~TeV and $2.5 < p_T < 4$~GeV (see\cite{9} for more details).
Below we will consider the phenomenological 
consequences for several LHC processes\footnote{The {\it MD'2018} gluon density will be implemented in
forthcoming release of the \textsc{tmdlib} package.}.

\section{Phenomenological applications} \indent

We are now in a position to apply the {\it MD'2018} gluon density 
to several hard processes studied at hadron colliders.
We use the $k_T$-factorization approach,
where the production cross section of any process under 
consideration (say, in $pp$ collisions) can be written as
\begin{equation}
  \displaystyle \sigma = \int dx_1 dx_2 \int d {\mathbf k}_{1T}^2 d {\mathbf k}_{2T}^2 \, f_{q/g}(x_1,{\mathbf k}_{1T}^2,\mu^2) f_{q/g}(x_2,{\mathbf k}_{2T}^2,\mu^2) \times \atop {
  \displaystyle  \times d\hat \sigma(x_1, x_2, {\mathbf k}_{1T}^2,{\mathbf k}_{2T}^2, \mu^2) },
\end{equation}

\noindent 
where $\hat \sigma(x_1, x_2, {\mathbf k}_{1T}^2,{\mathbf k}_{2T}^2, \mu^2)$ is the corresponding 
off-shell (depending on the transverse momenta of incoming particles) 
partonic cross section.
Everywhere below,
the multidimensional integration was performed
by the Monte Carlo technique, using the routine \textsc{vegas}\cite{52}.

\subsection{Proton structure functions $F_2^c$ and $F_2^b$} \indent

It is well-known that the basic information on the proton structure
can be extracted from deep inelastic $ep$
scattering. Its differential cross-section can be presented in the simple form:
\begin{equation}
  {d^2 \sigma \over dx dy} = {2 \pi \alpha^2 \over x Q^4} \left[ \left( 1 - y + {y^2\over 2}\right) F_2(x,Q^2) - {y^2 \over 2} F_L(x, Q^2)\right],
\end{equation}

\noindent
where $F_2(x, Q^2)$ and $F_L(x, Q^2)$ are the proton transverse and longitudinal structure functions, 
$x$ and $y$ are the usual Bjorken scaling variables.
In the present note we concentrate on the charm and beauty contributions
to $F_2(x, Q^2)$. The latter are described through perturbative generation of charm
and beauty quarks and, therefore, directly related with the gluon content of the proton.
Our evaluation below is based on the formulas\cite{53}.
Numerically, we apply the pole mass $m_c = 1.7$~GeV and $m_b = 4.8$~GeV
and strictly follow our previous consideration\cite{53} in all other aspects.

Our results are shown in Figs.~3 and~4 in comparison 
with the latest ZEUS\cite{54} and H1 data\cite{55,56}. The green and grey 
curves correspond to the predictions obtained with the {\it MD'2018} and 
{\it JH'2013 set 2} gluon densities, whereas the shaded bands 
represent the estimations of the scale uncertainties of these calculations.
We find that the {\it MD'2018} predictions for $F_2^c(x, Q^2)$ and $F_2^b(x, Q^2)$ 
are in reasonable agreement with the 
HERA data in a wide region of $x$ and $Q^2$, both in overall
normalization and shape. It slightly overshoots 
the {\it JH'2013 set 2} predictions at small $Q^2$ and low $x \leq 10^{-4}$.
At larger $Q^2$ and moderate and/or large $x \geq 10^{-2}$, the {\it JH'2013 set 2} 
gluon density function
tends to overestimate the HERA data on the structure function $F_2^b (x, Q^2)$,
that is due to the determination of input parameters of this gluon density at 
small $x$ only\cite{18}.
Therefore, the influence of the shape and other parameters
of the initial non-perturbative gluon distribution on the description
of the collider data is significant for a wide region of $x$
and $Q^2$\cite{9,10,11}.
The {\it MD'2018} gluon density, where all
these parameters are verified by the description of LHC data, 
leads to a better agreement with the HERA data, that confirms the 
link between soft processes at the LHC and low-$x$
physics at HERA, pointed out earlier\cite{9,10,11}.
Note that to estimate the scale uncertainties 
of the {\it JH'2013 set 2} calculations the method proposed in\cite{18} was used.
So, to evaluate the latter
we used the {\it JH'2013 set 2$+$} and {\it JH'2013 set 2$-$} sets instead
of default one {\it JH'2013 set 2}. These two sets 
represent a variation of the renormalization scale used in the 
off-shell production amplitude: the {\it JH'2013 set 2$+$} stands for a variation of $2\mu_R$, while
{\it JH'2013 set 2$-$} reflects $\mu_R/2$.
This method leads to somewhat reduced uncertainty bands in comparison
with the {\it MD'2018} predictions.

\subsection{Single top production at the LHC} \indent

Recently the CMS and ATLAS Collaborations have measured the differential cross sections of 
single top production (in the $t$-channel) at the LHC as a function of the transverse momenta 
and rapidity of the top quark and top-antiquark at $\sqrt s=8~$TeV\cite{57,58}.
Such measurements are known to be very useful for constraining parton densities in a proton\cite{59,60}.
To calculate the total and differential production 
cross sections we employ the four-flavor scheme (4FS), so that
the leading contribution comes from the $2\to3$ off-shell (reggeized) quark-gluon interaction subprocess:
\begin{equation}
q^*(k_1)+g^*(k_2)\to q^\prime(p_1)+\bar b(p_2)+t(p),
\end{equation}

\noindent
where the four-momenta of all particles are indicated in parentheses.
The main contribution to the amplitude~(8) comes from the diagram, 
which corresponds to initial gluon splitting to $b\bar b$ pair with subsequent
exchange of $W$-boson between the $b$ and the light quark. The latter reads
\begin{equation}
  \displaystyle \mathcal A = -g\frac{e^2}{8\sin^2\theta_W}V_{qq^\prime}V_{tb}\bar u_{s_1}(p_1)\Gamma^{\mu}_{(+)}(k_1,-p_1)(1-\gamma^5) u_{s_2}(x_1l_1)\times \atop{
  \displaystyle \bar u_{s_3}(p)\gamma_{\mu}(1-\gamma^5)\frac{\hat k_2-\hat p_2+m_b}{(k_2-p_2)^2-m_b^2}\hat\epsilon(k_2)v_{s_4}(p_2)t^a\frac{1}{(p_1 - k_1)^2-m_W^2+im_W\Gamma_W}},
\end{equation}

\noindent
where $g$ and $e$ are the strong and electric charges respectively, $\theta_W$ is the weak Weinberg mixing angle, 
$V_{q_aq_b}$ are the Cabibbo-Kobayashi-Maskawa (CKM) matrix elements, $m_b$ and $m_W$ are the $b$-quark 
and $W$-boson masses, $a$ is the eight-fold color index and $\Gamma_W$ is the $W$-boson full decay width.
The effective vertex $\Gamma^{\mu}_{(+)}(k,q)$, 
that ensures gauge invariance of the amplitude~(9)
despite the off-shell initial partons, can be written as\cite{61,62}:
\begin{equation}
\Gamma^{\mu}_{(+)}(k,q) = \gamma^\mu-\hat k\frac{l^\mu_1}{l_1\cdot q},
\end{equation}

\noindent
where $l_1$ is the proton four-momentum ($k_1 = x_1 l_1 + k_{1T}$ and $k_2 = x_2 l_2 + k_{2T}$).
The polarization sum for the off-shell gluon is taken in the BFKL form\cite{3,4}:
\begin{equation}
\sum\epsilon^\mu(k)\epsilon^\nu(k) = \frac{k_T^\mu k_T^\nu}{\mathbf k_T^2}.
\end{equation}

\noindent
In all other aspects the calculation is straightforward and follows standard Feynman rules. 
The evaluation of traces was performed using the algebraic manipulation system \textsc{form}.

Having calculated the squared amplitude~(9), one can evaluate the total and differential cross sections
of single top production according to the TMD factorization formula~(7).
Numerically, we took $m_W = 80.4$~GeV and $\Gamma_W=2.1$~GeV. The light quarks were kept massless, 
while for heavy quarks we took $m_b = 4.75$~GeV and $m_t = 175$~GeV.
The weak mixing angle was chosen to correspond to $\sin^2\theta_W=0.23$\cite{63}.
As the renormalization $\mu_R$ and factorization $\mu_F$ scales, we choose the largest mass parameter in our 
calculation, the top tranverse mass\footnote{The different choices of hard scales in the single top production 
are discussed in\cite{60,64}.}.

The results of our calculations for single top quark production in $t$-channel 
are presented in Fig.~5 --- 7 in comparison 
with the CMS and ATLAS data\cite{57,58}. 
These data correspond to the absolute and normalized cross-sections on parton level, 
differential on top quark transverse momentum and rapidity. 
Studying the latter could lead to a more stringent 
comparison between data and theory due to reduced experimental (mainly systematic)
and theoretical (scale) uncertainties.
We find that both the {\it MD'2018} and {\it JH'2013 set 2} gluon densities predict almost identical normalized cross sections, 
which agree with the CMS and ATLAS measurement quite well. However, the {\it MD'2018} density results in 
a little smaller total cross section than the {\it JH'2013 set 2} one, that leads to somewhat better 
description of the data. 
Thus the calculations endorse the usage of the {\it MD'2018} gluon density
for evaluation of cross sections of processes with quite large $x$ values involved.
Note that the size of scale uncertainties of {\it MD'2018} and {\it JH'2013 set 2} calculations
are rather close to each other in the kinematical region probed.



\subsection{Inclusive Higgs boson production at the LHC} \indent

Very recently the CMS and ATLAS Collaborations have reported measurements\cite{65,66,67,68}
of the total and differential cross sections of inclusive Higgs boson production
at $\sqrt s = 13$~TeV obtained in different Higgs decay channels.
These measurements can be used to investigate the gluon dynamics in a proton
since the dominant mechanism of inclusive Higgs production at the LHC is 
gluon-gluon fusion\cite{69,70,71,72}.
Here, to calculate the total and differential cross sections of Higgs boson
production we strictly follow our previous consideration\cite{73}.
The latter is based on the off-shell amplitude of the gluon-gluon fusion 
subprocess $g^* g^* \to H$ calculated using
the effective Lagrangian\cite{74,75} for the Higgs coupling 
to gluons and extended recently
to the subsequent $H\to \gamma \gamma$, $H \to W^+W^- \to e^\pm \mu^\mp \nu \bar \nu$\cite{76} and 
$H \to ZZ^* \to 4l$ decays\cite{76,77}.
The details of the calculations are explained in\cite{76}.
Below we present the numerical results obtained with the 
{\it MD'2018} and {\it JH'2013 set 2} gluon densities 
for $H \to \gamma \gamma$ and $H \to ZZ^* \to 4l$ decay modes.

The CMS and ATLAS measurements refer to a restricted part of the phase space (fiducial phase
space)  defined  to  match  the  experimental  acceptance  in  terms  of  the  photon  kinematics
and event selection.
In the CMS analysis\cite{65} two isolated photons originating from the Higgs boson decays 
are required to have pseudorapidities $|\eta^\gamma| < 2.5$. 
Photons with largest and next-to-largest transverse momentum $p_T^\gamma$
(so-called leading and subleading photons) must satisfy the 
conditions of $p_T^\gamma/m^{\gamma \gamma} > 1/3$ and $p_T^\gamma/m^{\gamma \gamma} > 1/4$
respectively, where the diphoton mass $m^{\gamma \gamma}$ is required to be 
$100 < m^{\gamma \gamma} < 180$~GeV. In the ATLAS measurement\cite{67} both decay photons 
must have pseudorapidities $|\eta^\gamma| < 2.37$
with the leading (subleading) photon satisfying $p_T^\gamma/m^{\gamma \gamma} > 0.35$~($0.25$), 
while invariant mass $m^{\gamma \gamma}$
is required to be $105 < m^{\gamma \gamma} < 160$~GeV.
In the $H \to ZZ^* \to 4l$ decay channel, only events with a four-lepton
invariant mass $118 < m_{4l} < 129$~GeV are kept by ATLAS Collaboration\cite{68} and  
each lepton (electron or muon) must satisfy transverse momentum cut
$p_T > 6$~GeV and be in the pseudorapidity range $|\eta| < 2.47$.
The highest-$p_T$ lepton
in the quadruplet must have $p_T > 20$~GeV and the second (third) lepton in $p_T$ order
must satisfy $p_T > 15$($10$)~GeV. These leptons are required to be separated from each other
by $\Delta R = \sqrt{(\Delta \eta)^2 + (\Delta \phi)^2} > 0.1$($0.2$) when having the same (different)
lepton flavors. The invariant mass $m_{12}$ of the lepton pair closest to the $Z$ boson mass (leading pair)
is required to be $50 < m_{12} < 106$~GeV.
The subleading pair is chosen as the remaining lepton pair with invariant mass $m_{34}$ closest to the
$Z$ boson mass and satisfying the requirement $12 < m_{34} < 115$~GeV.
The CMS measurement\cite{66} requires at least four leptons in the event with at least
one lepton having $p_T > 20$~GeV, another lepton having $p_T > 10$~GeV and the remaining ones
having $p_T > 7$ and $5$~GeV respectively. All leptons must have the 
pseudorapidity $|\eta| < 2.4$, the leading pair invariant mass $m_{12}$ must be
$40 < m_{12} < 120$~GeV and subleading one should be $12 < m_{34} < 120$~GeV.
Finally, the four-lepton invariant mass $m_{4l}$ must satisfy $105 < m_{4l} < 140$~GeV cut.

The results of our calculations are shown in Figs.~8 and~9 
in comparison with latest LHC data\cite{65,66,67,68}.
In the $H \to \gamma \gamma$ decay channel, 
we calculated the distributions on the diphoton pair 
transverse momentum $p_T^{\gamma \gamma}$, absolute value of 
the rapidity $|y^{\gamma \gamma}|$, photon helicity
angle $|\cos \theta^*|$ (in  the  Collins-Soper  frame) and 
azimuthal angle difference $\Delta \phi^{\gamma \gamma}$
between the produced photons. 
In the $H \to ZZ^* \to 4l$ decay channel, we calculated distributions on the Higgs
transverse  momentum $p_T^H$, rapidity $|y^H|$, invariant mass of the subleading lepton 
pair $m_{34}$ and cosine of the leading lepton pair decay angle $|\cos \theta^*|$
in the four-lepton rest frame with respect to the beam axis.
We find that both the {\it MD'2018} and {\it JH'2013 set 2}
predictions reasonably agree with the data within the uncertainties 
for all considered kinematical observables,
although the {\it MD'2018} results lie a bit below {\it JH'2013 set 2} ones.
Some tendency to underestimate the LHC data at large 
transverse  momenta could be explained by the missing contributions 
from the weak  boson fusion ($W^+W^- \to H$ and $ZZ \to H$)
and/or associated $HZ$ or $HW^{\pm}$ production\cite{78}, 
which become important at high $p_T$
and are not taken into account in our present consideration.
The measured rapidity, $|\cos \theta^*|$ and $m_{34}$ distributions
are well reproduced by our calculations.
As one can see, despite the fact that both the {\it MD'2018} and {\it JH'2013 set 2} gluon distributions
agree with the available data, the inclusive Higgs boson 
production at the LHC is very sensitive to the TMD gluon density in a proton, in particular, to
the parameters of the initial TMD gluon distribution. It could be important to further
constrain the latter.

\section{Conclusion} \indent

We have refined a fit of the experimental data on the inclusive spectra of the 
charged particles produced in the central
$pp$ and $AA$ collisions at RHIC and the LHC to determine the TMD gluon density in
a proton at the starting scale.
The parameters of this fit do not depend on the
initial energy in a wide energy interval.
Using a numerical solution of the CCFM gluon evolution equation,
we extended the derived TMD gluon
density (denoted as {\it Moscow-Dubna 2018}, or {\it MD'2018} set) 
to a whole kinematical region
and supplied it with the relevant TMD valence 
and sea quark distributions.
The latter was calculated in the approximation where the 
gluon-to-quark splitting occured at the last evolution step 
using the TMD gluon-to-quark splitting function.
Some phenomenological applications of the proposed {\it MD'2018} 
quark and gluon densities to the hard LHC processes were discussed.
We demonstrated a significant influence of the initial 
non-perturbative gluon distribution on the description of the LHC data, that
is important to further precise determination 
of the TMD quark and gluon densites in a proton.

\section{Acknowledgements} \indent

We would like to thank S.P.~Baranov and F.~Hautmann for very useful 
discussions and important remarks.
A.V.L. and M.A.M. are grateful to DESY Directorate for the 
support in the framework of Moscow --- DESY project on Monte-Carlo implementation for
HERA --- LHC. M.A.M. was also supported by a grant of the foundation for
the advancement of theoretical physics and mathematics "Basis" 17-14-455-1.
Part of this work was done by M.A.M. during his stay at DESY, 
funded by DAAD (Program "Research Stays for University Academics and Scientists").

\newpage

\begin{figure}
\begin{center}
\includegraphics[width=7.9cm]{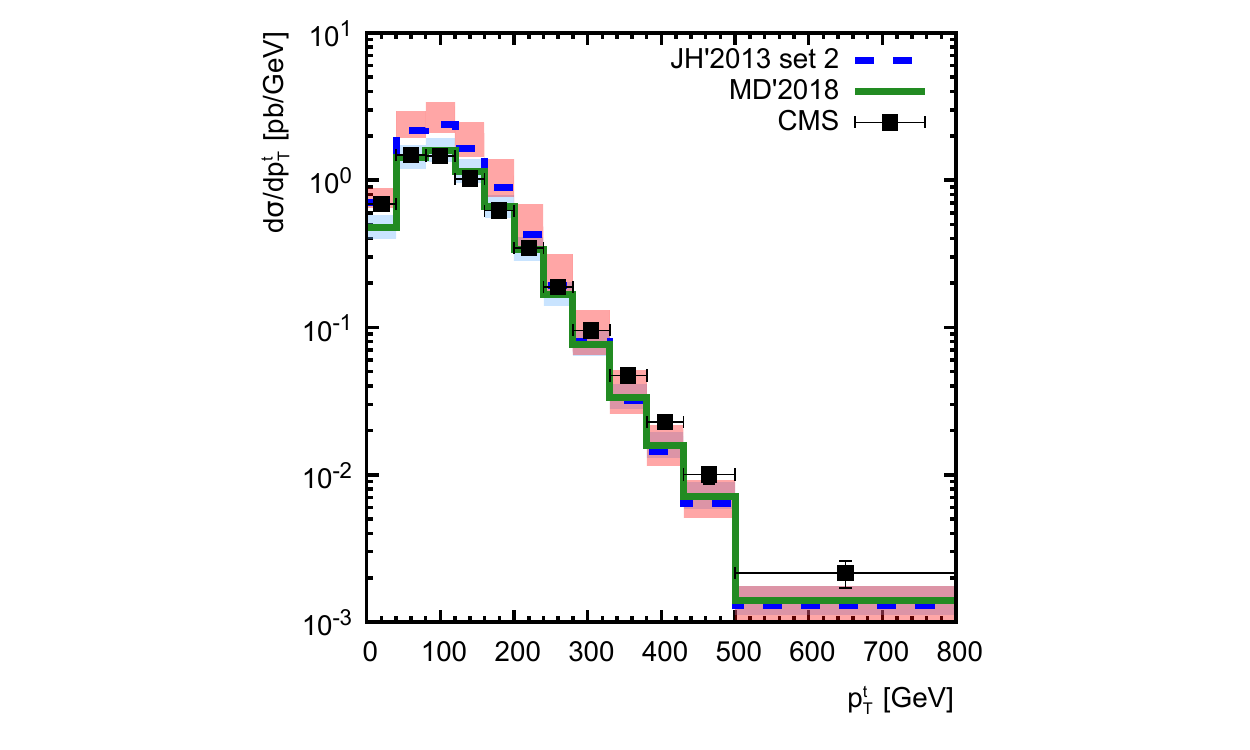}
\includegraphics[width=7.9cm]{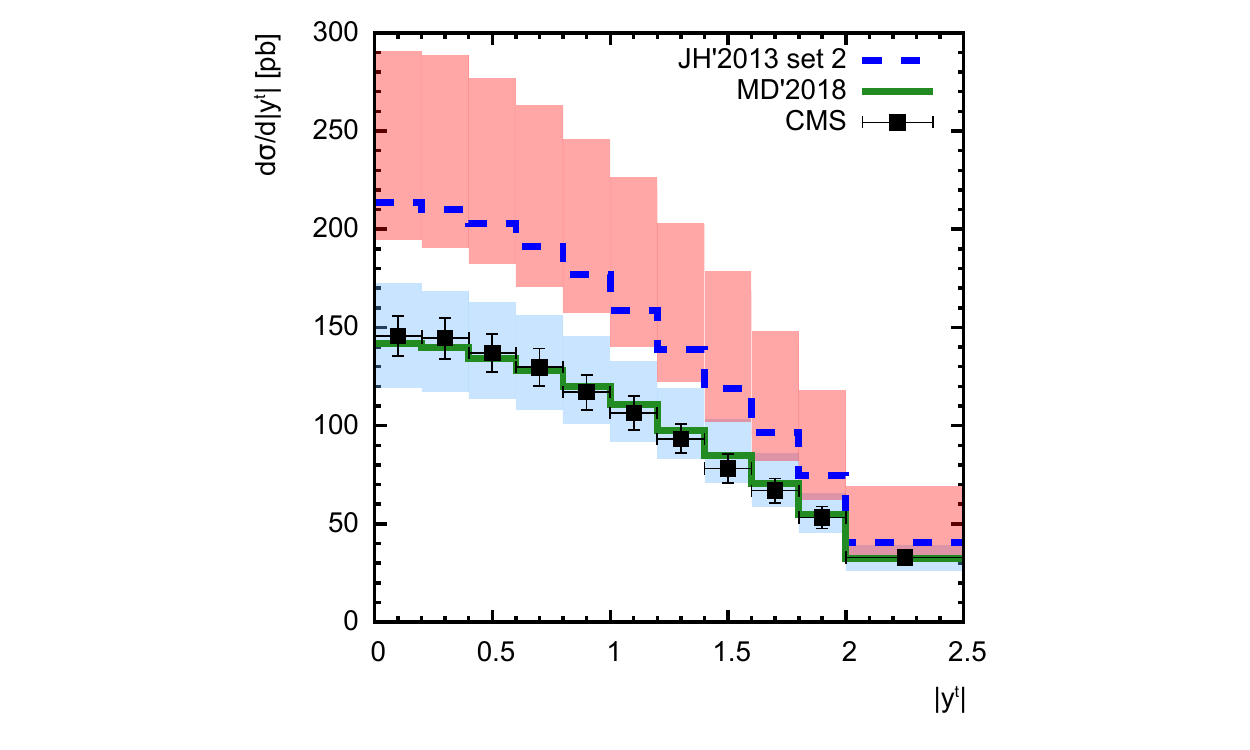}
\caption{The transverse momentum and rapidity distributions of inclusive $t\bar t$ production in
$pp$ collisions at $\sqrt s = 13$~TeV. 
The green (solid) and blue (dashed) curves correspond to the predictions obtained using the 
{\it MD'2018} and {\it JH'2013 set 2} gluons, respectively. 
The shaded bands represent their scale uncertainties.
The experimental data are from CMS\cite{34}.}
\label{fig2}
\end{center}
\end{figure}

\begin{figure}
\begin{center}
\includegraphics[width=7.9cm]{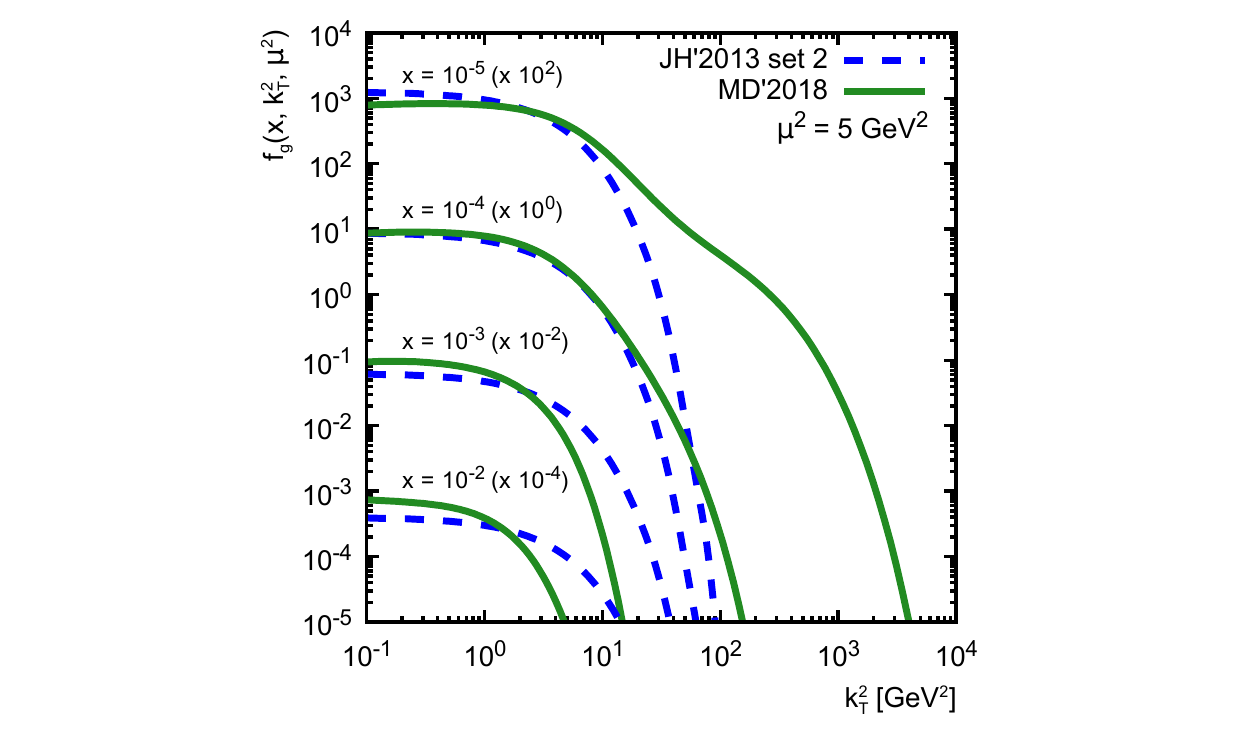}
\includegraphics[width=7.9cm]{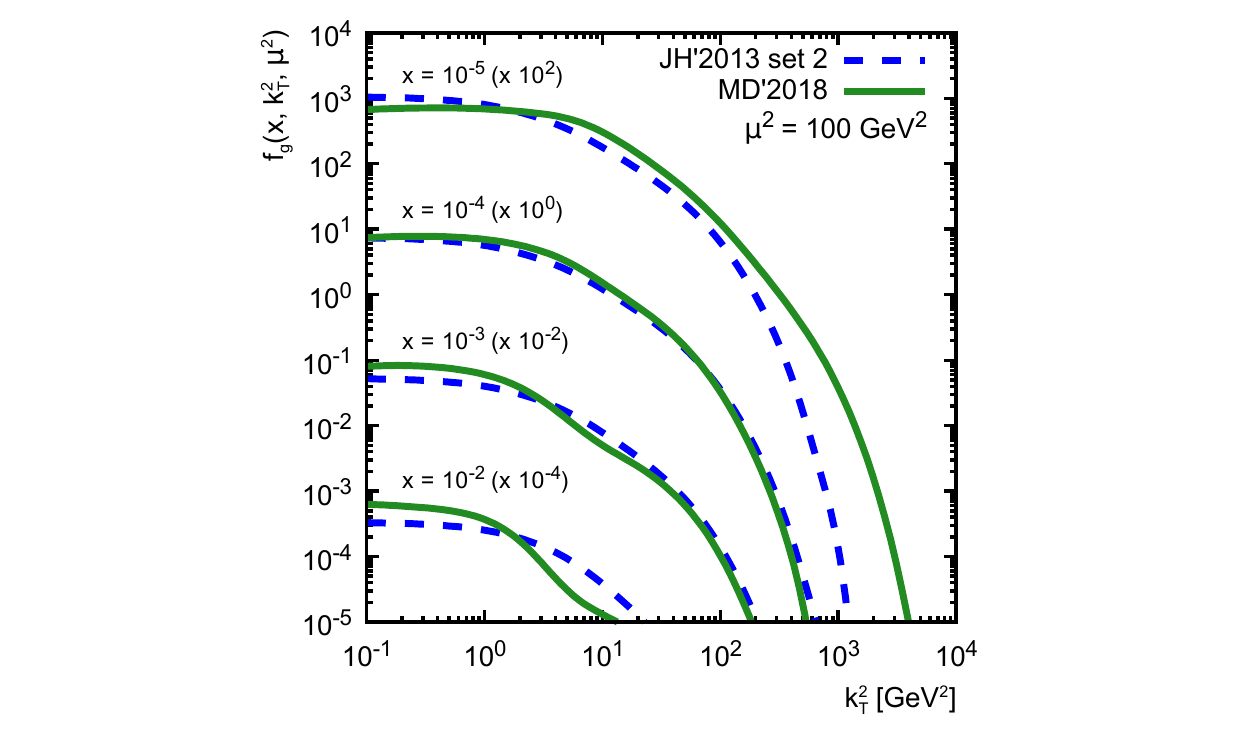}
\caption{The TMD gluon densities in the proton calculated as a function
of the gluon transverse momentum ${\mathbf k}_T^2$ at different longitudinal momentum 
fraction $x$ and $\mu^2$ values. The green (solid) and blue (dashed) curves correspond to the 
{\it MD'2018} and {\it JH'2013 set 2} gluon density functions, respectively.}
\label{fig3}
\end{center}
\end{figure}

\begin{figure}
\begin{center}
\includegraphics[width=15cm]{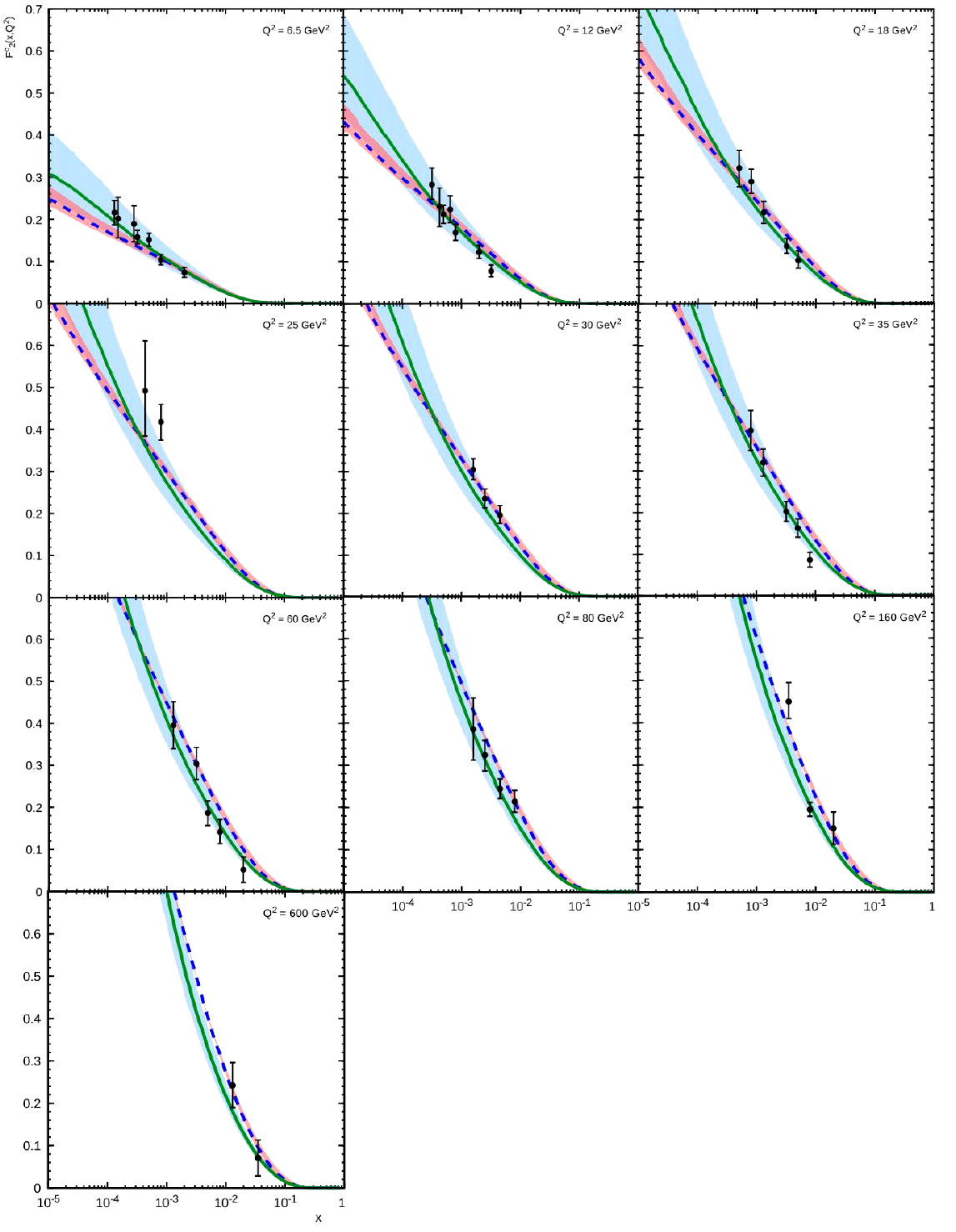}
\caption{The charm contribution to the proton structure function $F_2(x, Q^2)$
as a function of $x$ calculated at different $Q^2$. 
Notation of histograms is the same as in Fig.~2.
The experimental data are from ZEUS\cite{54} and H1\cite{55}.}
\label{fig4}
\end{center}
\end{figure}

\begin{figure}
\begin{center}
\includegraphics[width=15cm]{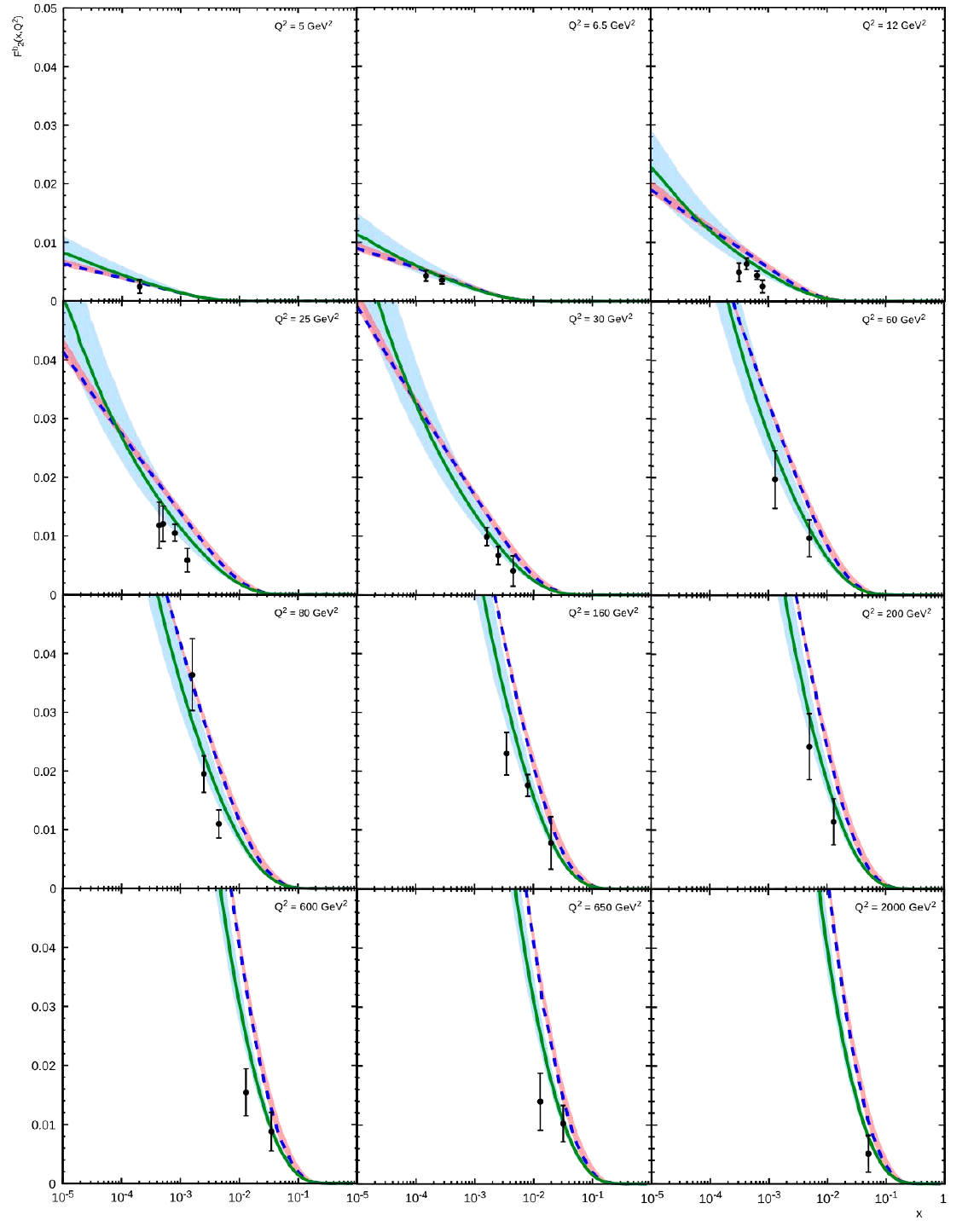}
\caption{The beauty contribution to the proton structure function $F_2(x, Q^2)$
as a function of $x$ calculated at different $Q^2$. Notation of histograms is the same as in Fig.~2.
The experimental data are from ZEUS\cite{54} and H1\cite{56}.}
\label{fig5}
\end{center}
\end{figure}

\begin{figure}
\begin{center}
\includegraphics[width=7.9cm]{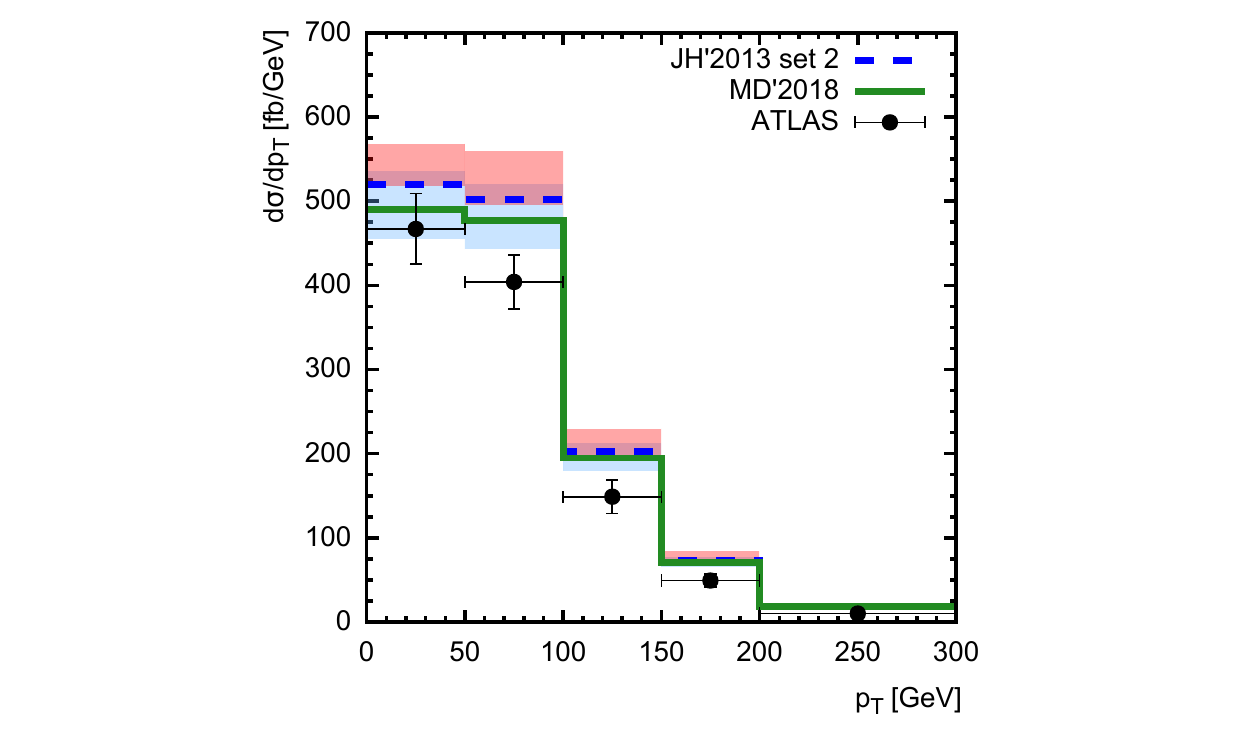}
\includegraphics[width=7.9cm]{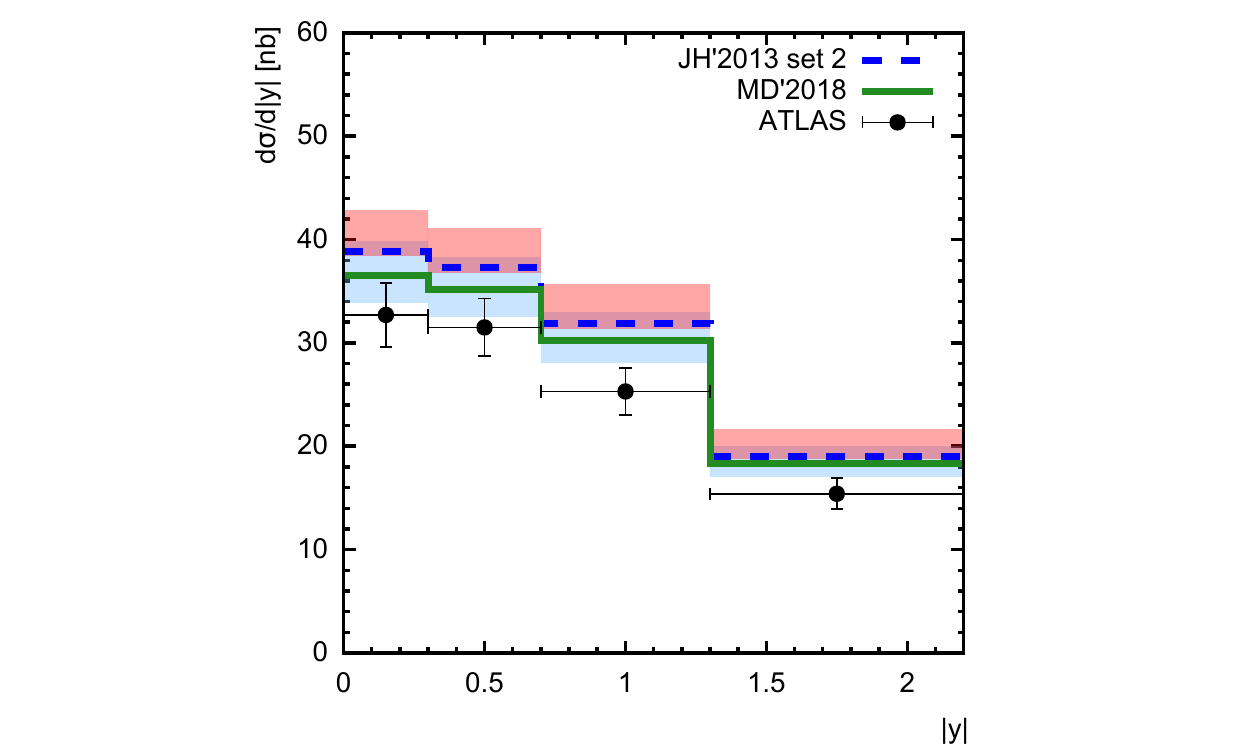}
\includegraphics[width=7.9cm]{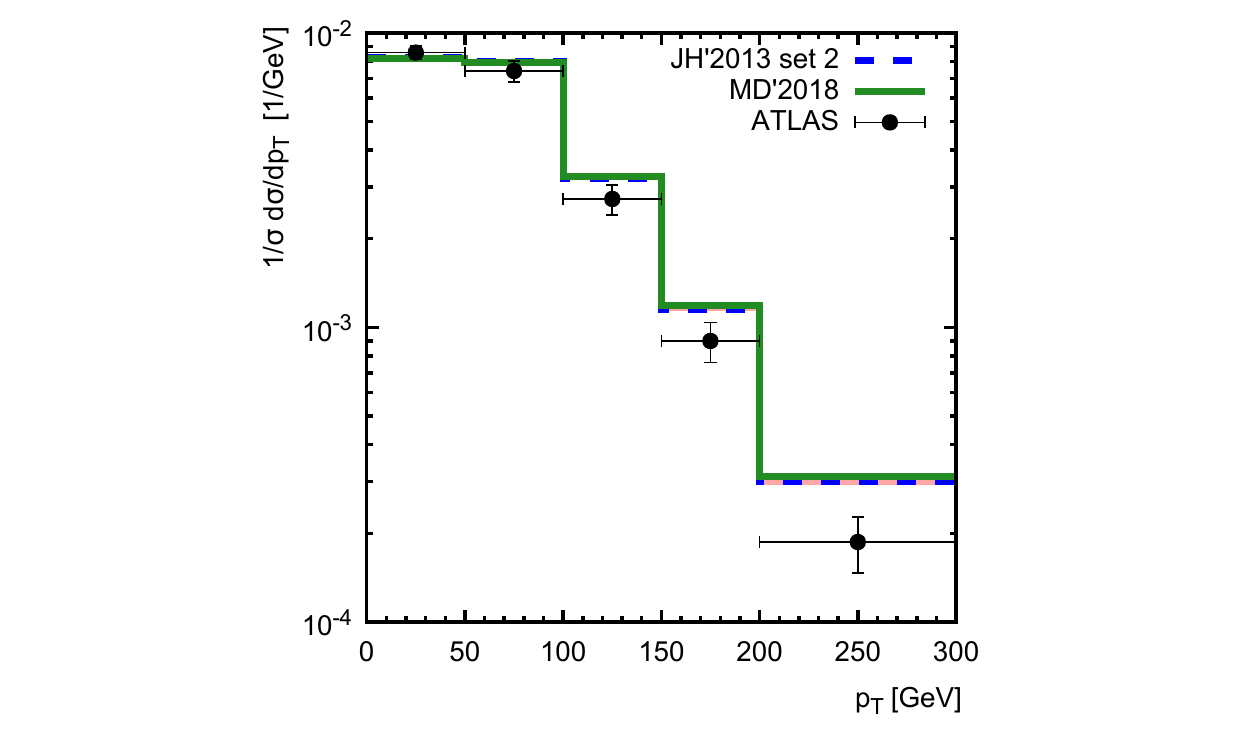}
\includegraphics[width=7.9cm]{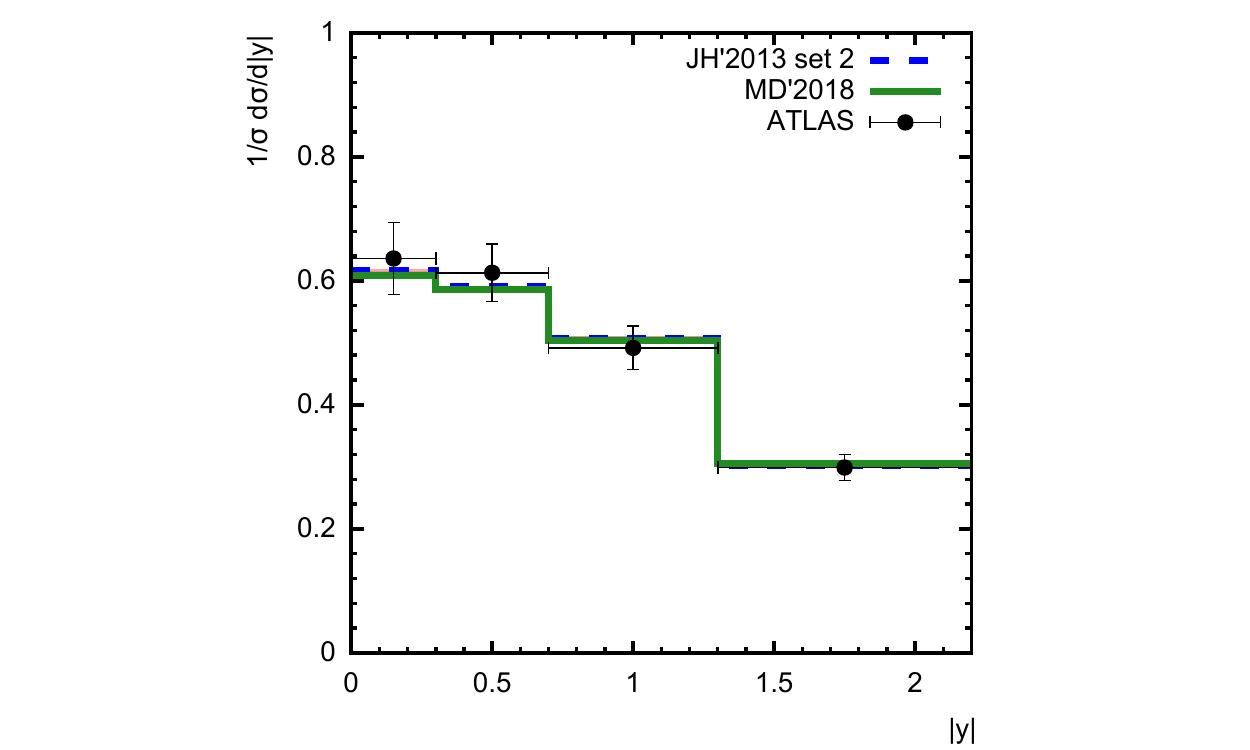}
\caption{The differential cross sections of inclusive $t$-channel single top production 
at $\sqrt s = 8$~TeV as a functions of top quark transverse momentum and rapidity. 
Notation of histograms is the same as in Fig.~2.
The experimental data are from ATLAS\cite{58}.}
\label{fig6}
\end{center}
\end{figure}

\begin{figure}
\begin{center}
\includegraphics[width=7.9cm]{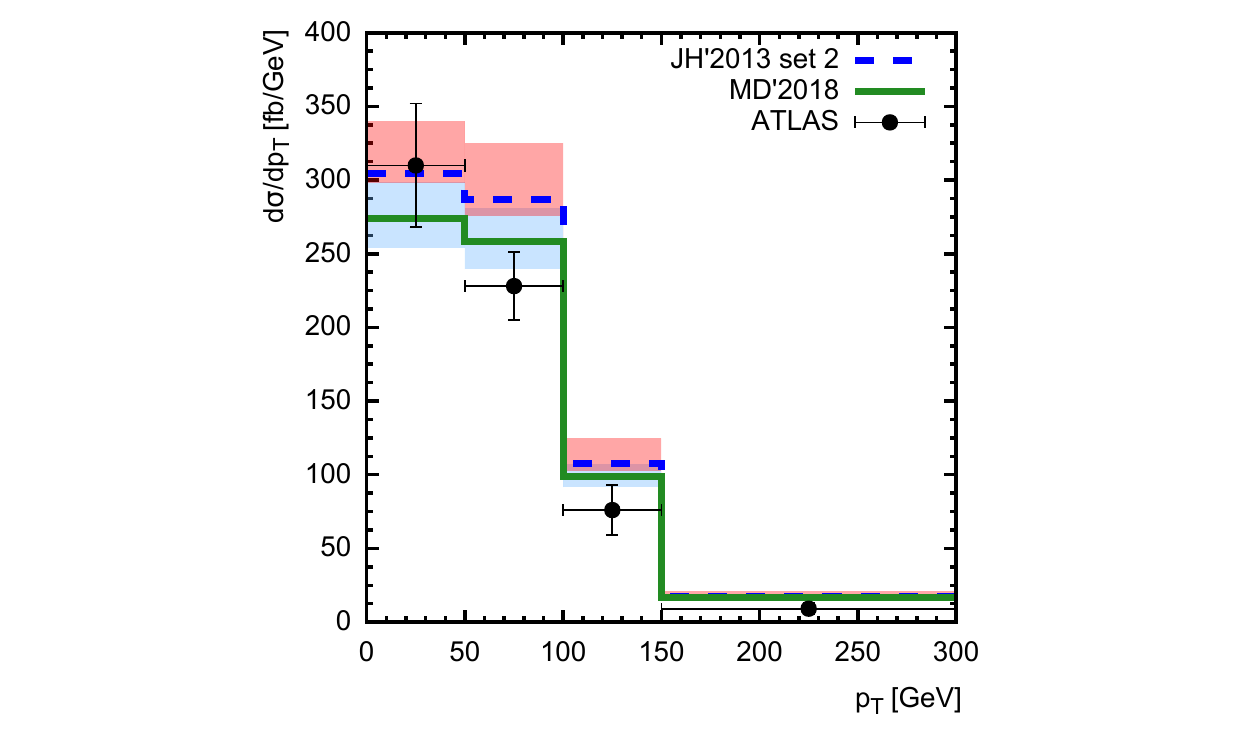}
\includegraphics[width=7.9cm]{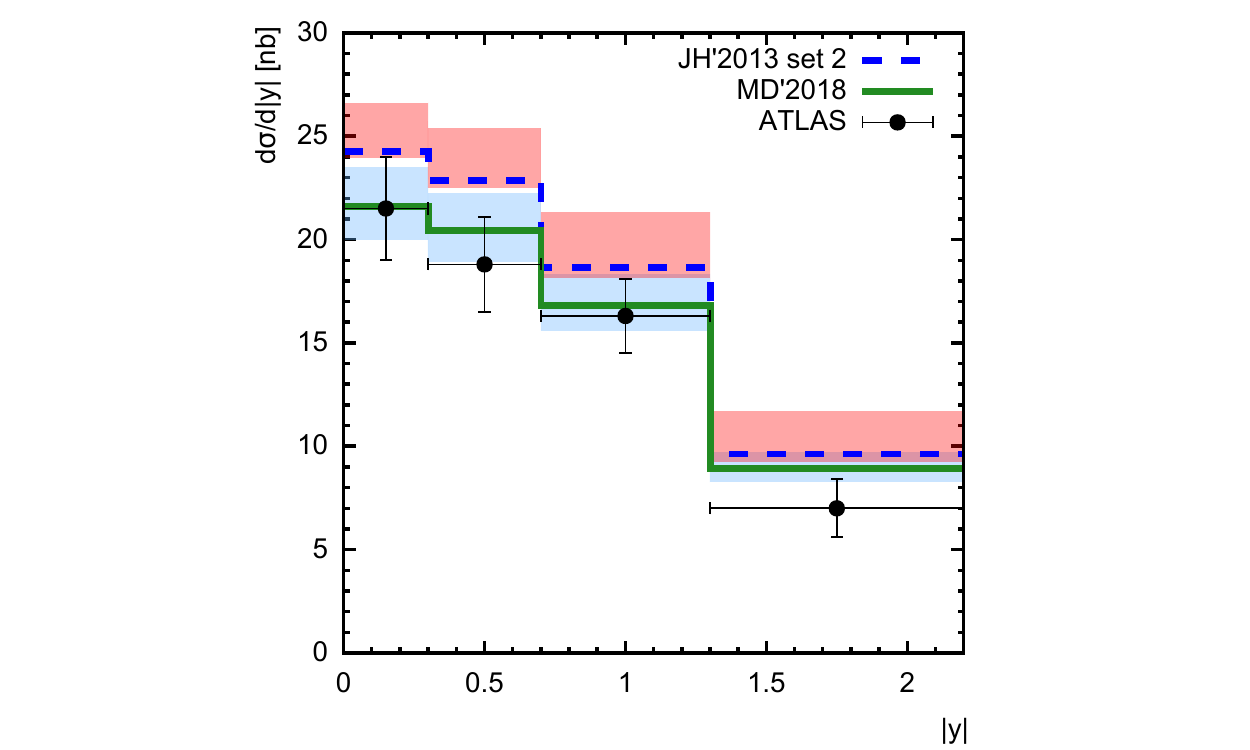}
\includegraphics[width=7.9cm]{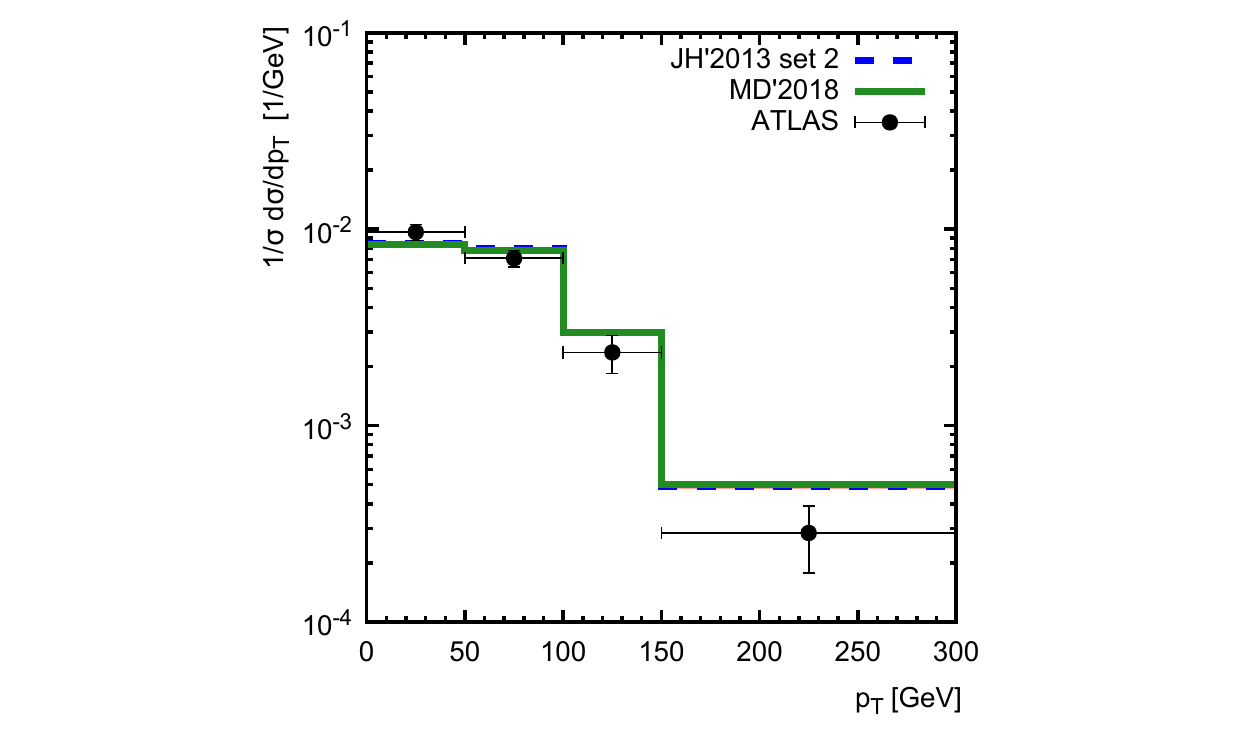}
\includegraphics[width=7.9cm]{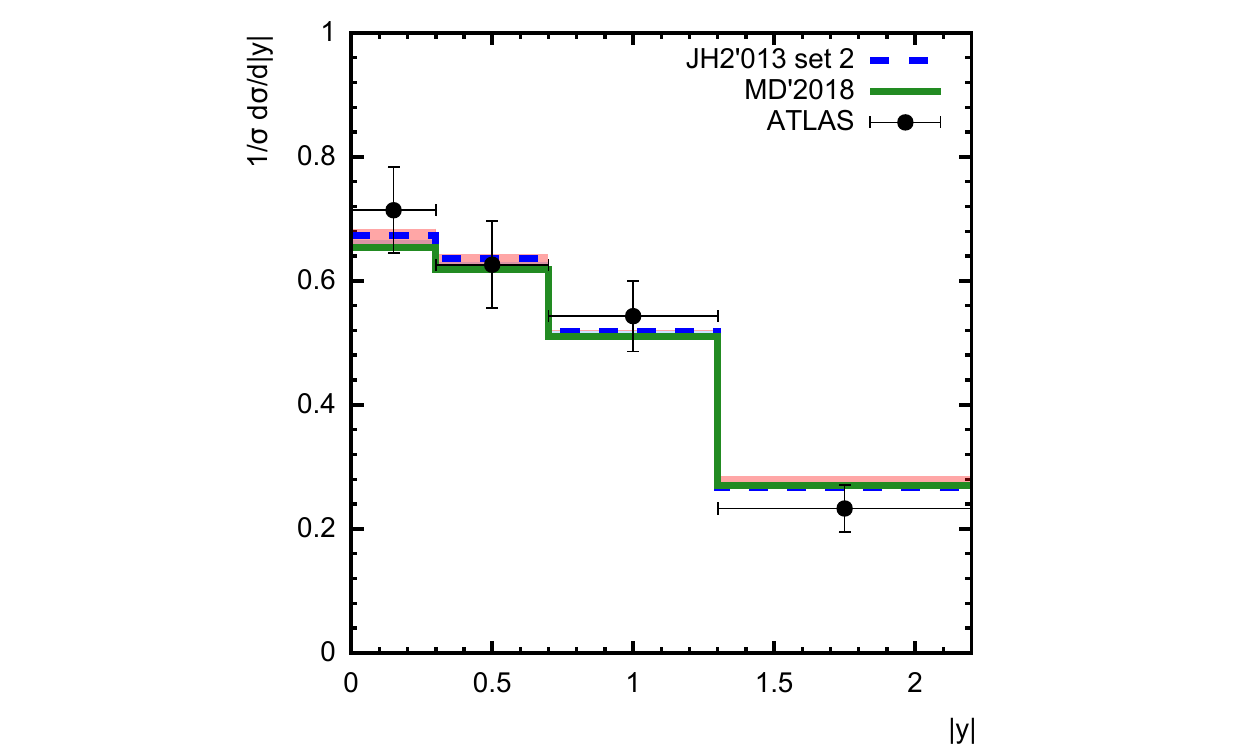}
\caption{The differential cross sections of inclusive $t$-channel single anti-top production 
at $\sqrt s = 8$~TeV as a functions of $\bar t$ quark transverse momentum and rapidity.
Notation of histograms is the same as in Fig.~2.  
The experimental data are from ATLAS\cite{58}.}
\label{fig7}
\end{center}
\end{figure}

\begin{figure}
\begin{center}
\includegraphics[width=7.9cm]{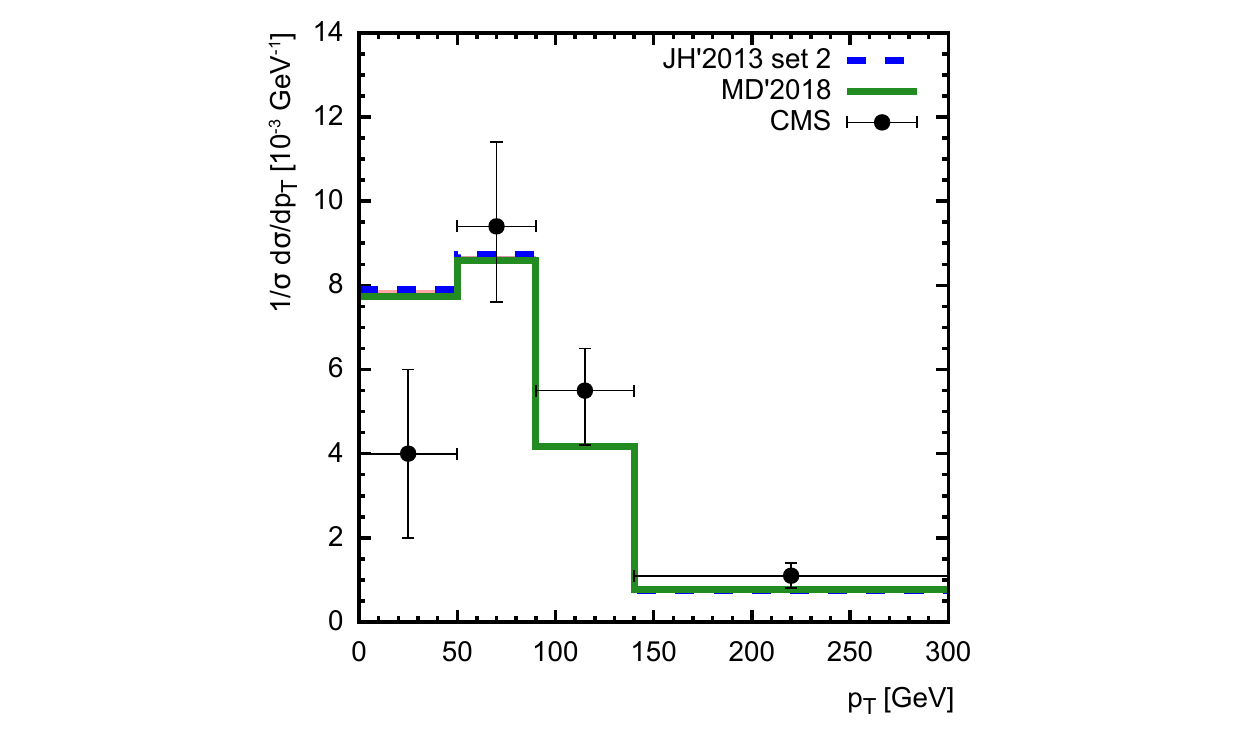}
\includegraphics[width=7.9cm]{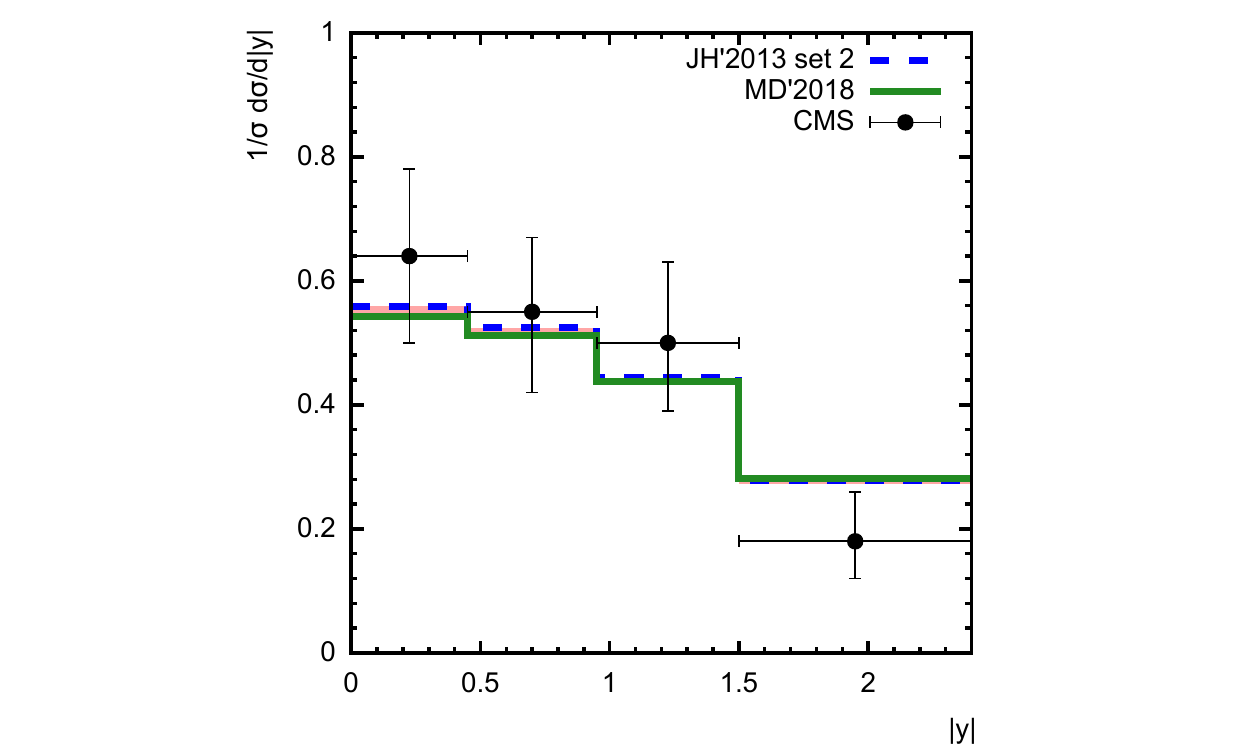}
\caption{The normalized differential cross sections of inclusive $t$-channel single top production 
at $\sqrt s = 8$~TeV as a functions of top quark transverse momentum and rapidity.
Notation of histograms is the same as in Fig.~2.  
The experimental data are from CMS\cite{57}.}
\label{fig8}
\end{center}
\end{figure}

\begin{figure}
\begin{center}
\includegraphics[width=7.9cm]{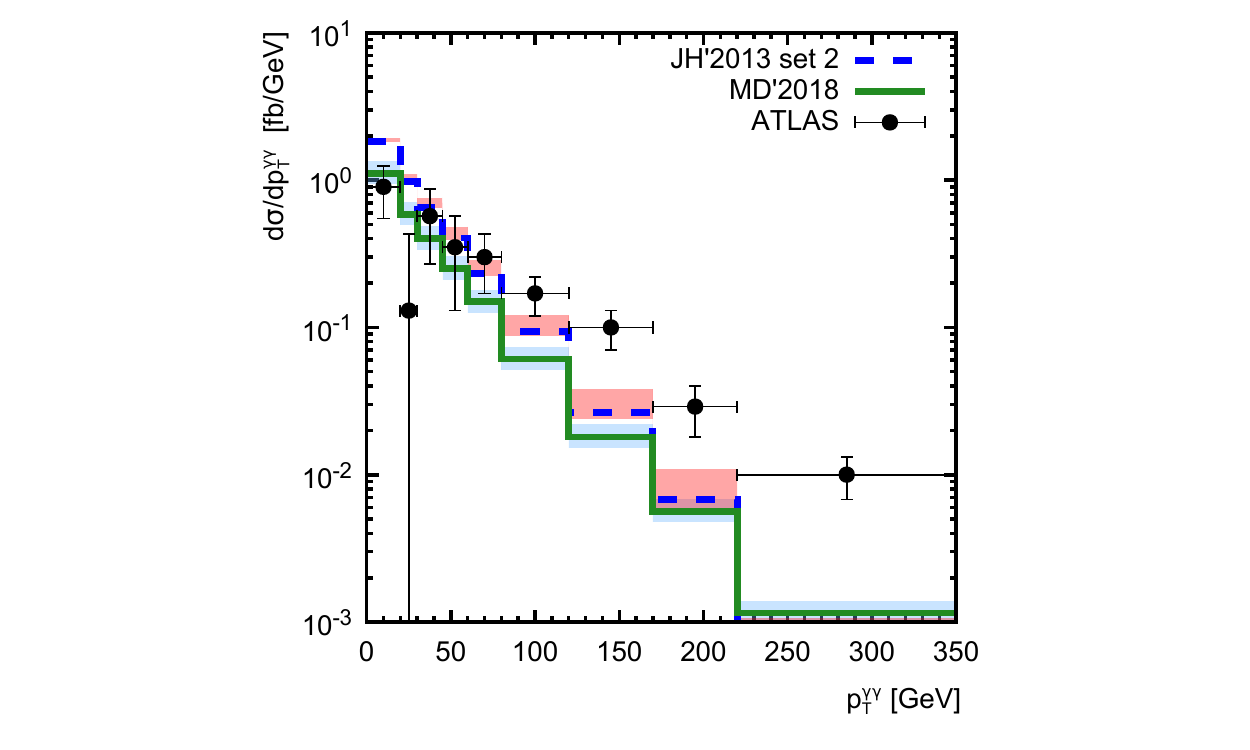}
\includegraphics[width=7.9cm]{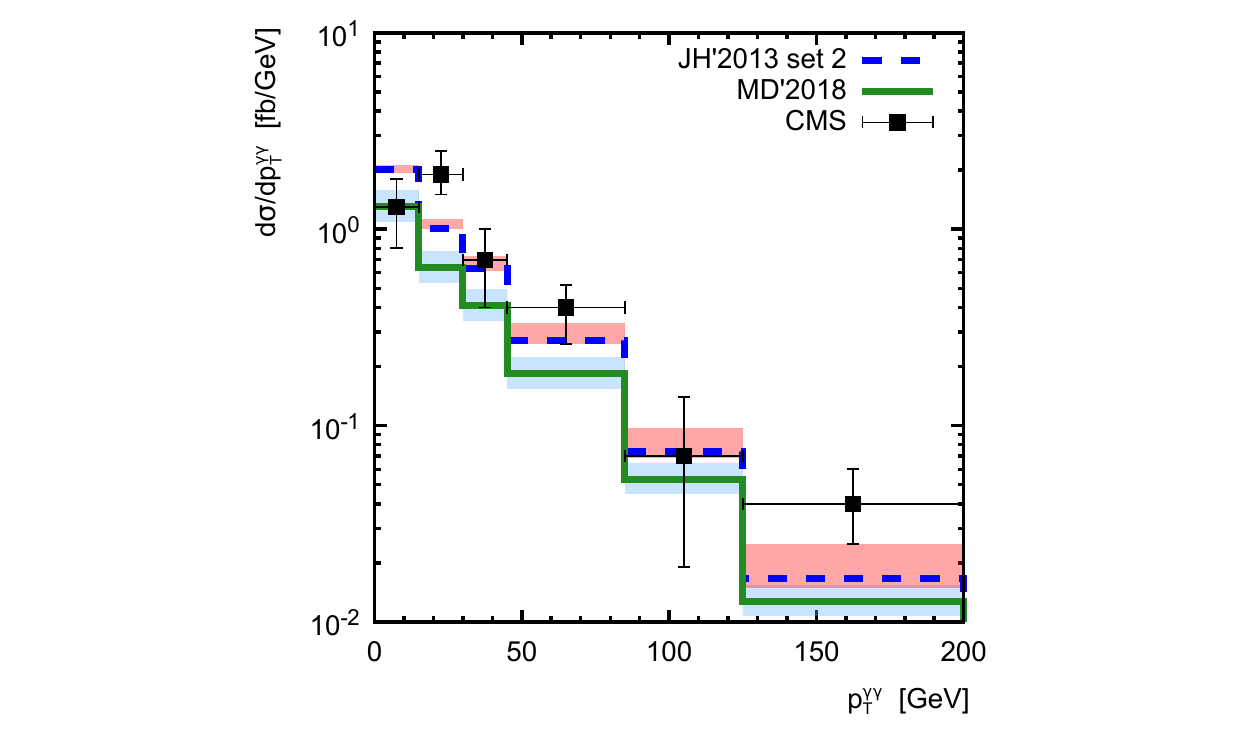}
\includegraphics[width=7.9cm]{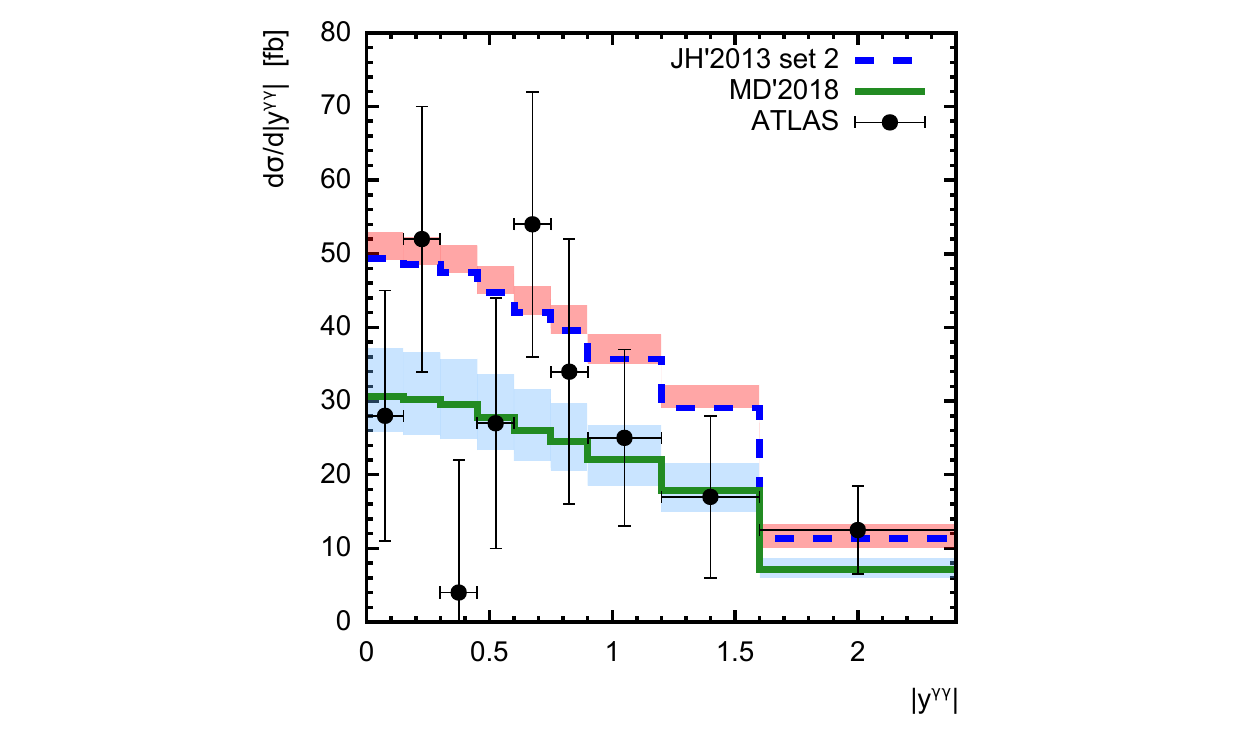}
\includegraphics[width=7.9cm]{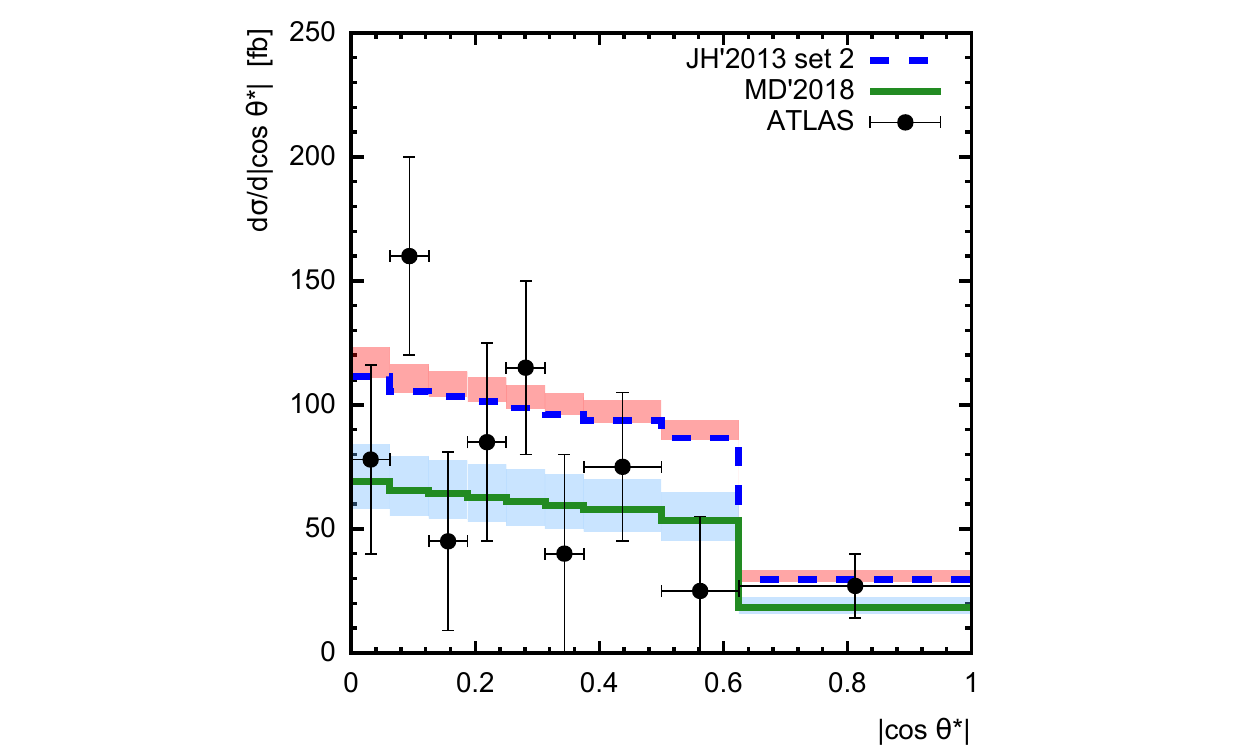}
\caption{The differential cross sections of inclusive Higgs boson 
production (in the diphoton decay mode) at $\sqrt s = 13$~TeV
as functions of diphoton pair transverse momentum $p_T^{\gamma \gamma}$, rapidity $|y^{\gamma \gamma}|$ 
and photon helicity angle $\cos \theta^*$ in the Collins-Soper frame. 
Notation of histograms is the same as in Fig.~2. 
The experimental data are from CMS\cite{65} and ATLAS\cite{67}.}
\label{fig10}
\end{center}
\end{figure}

\begin{figure}
\begin{center}
\includegraphics[width=7.9cm]{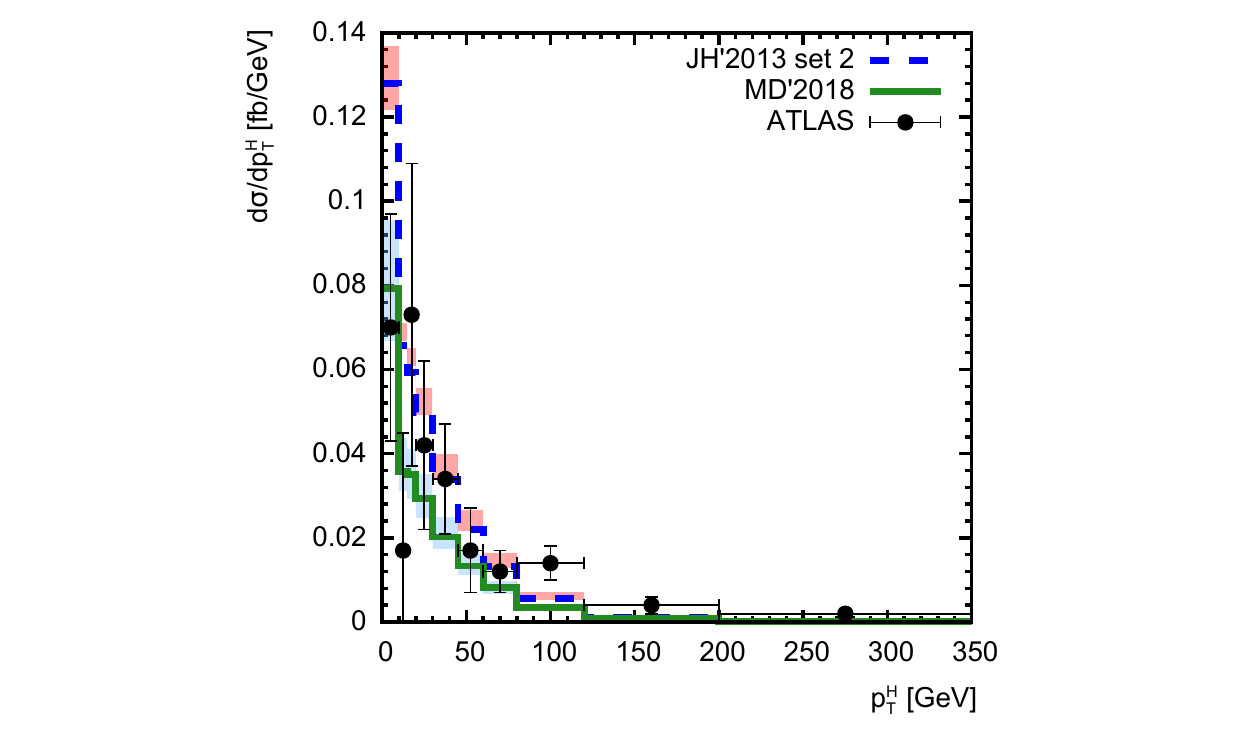}
\includegraphics[width=7.9cm]{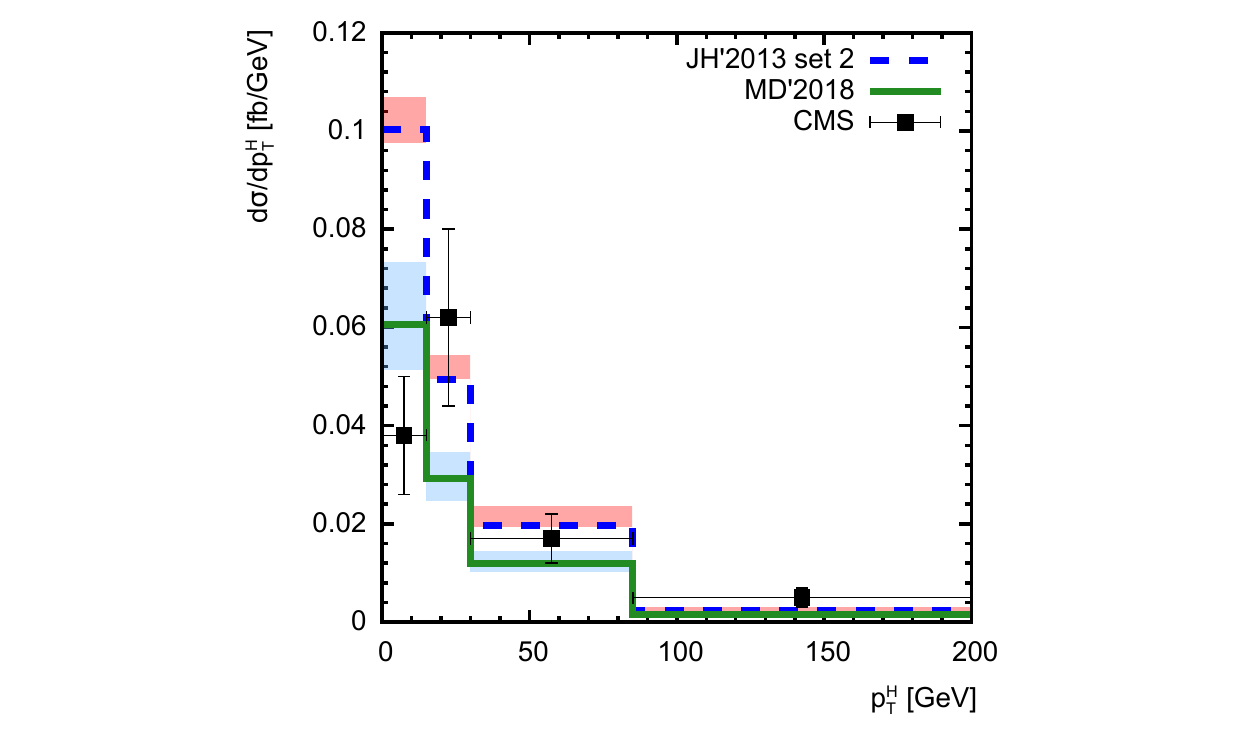}
\includegraphics[width=7.9cm]{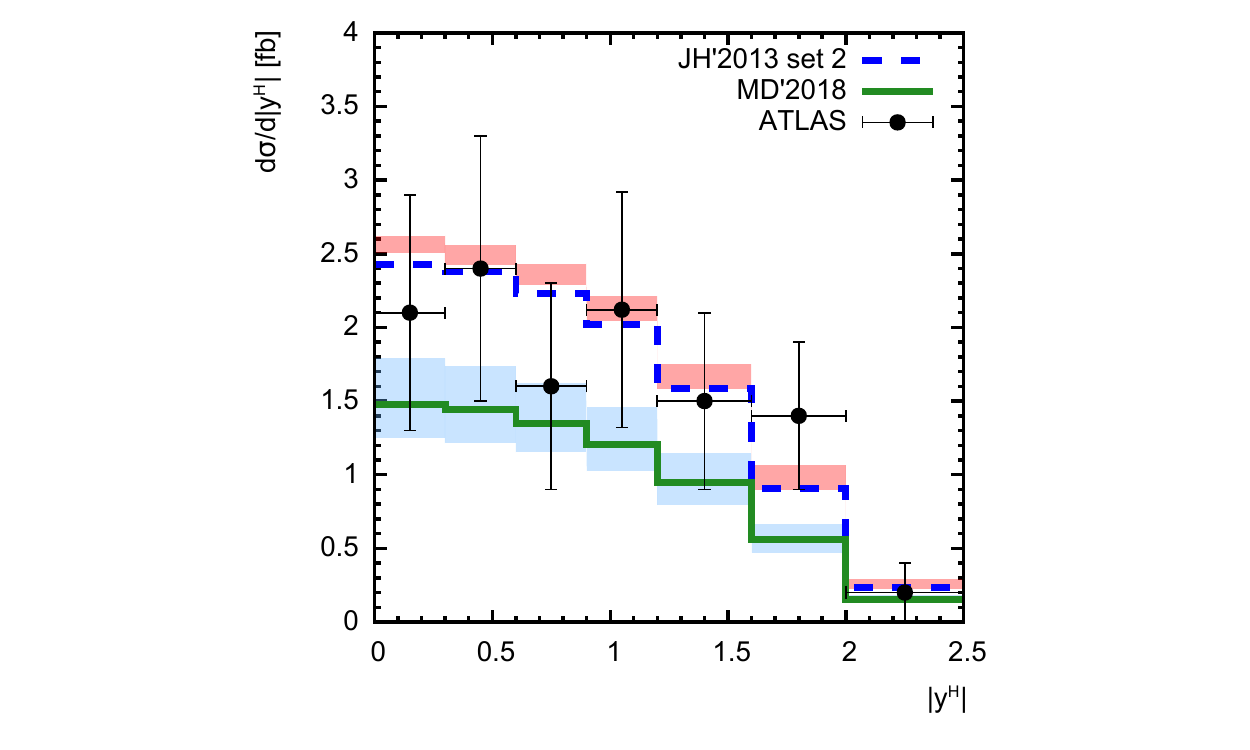}
\includegraphics[width=7.9cm]{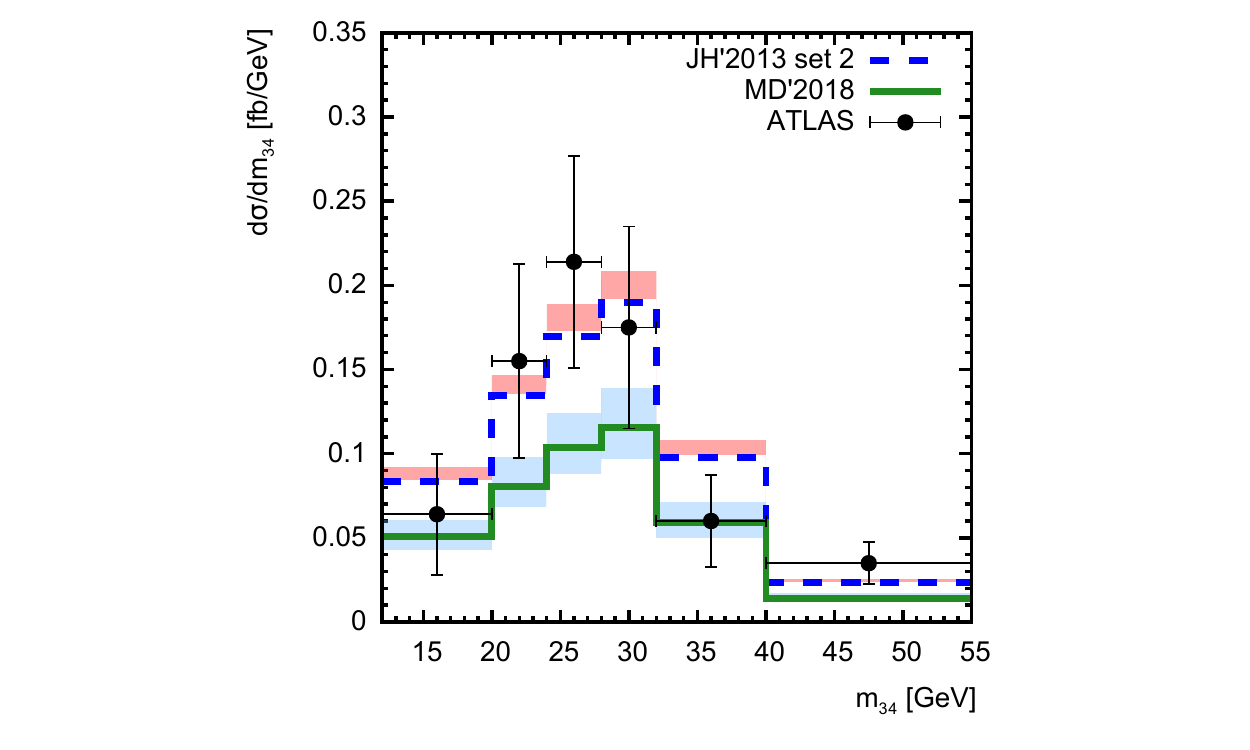}
\includegraphics[width=7.9cm]{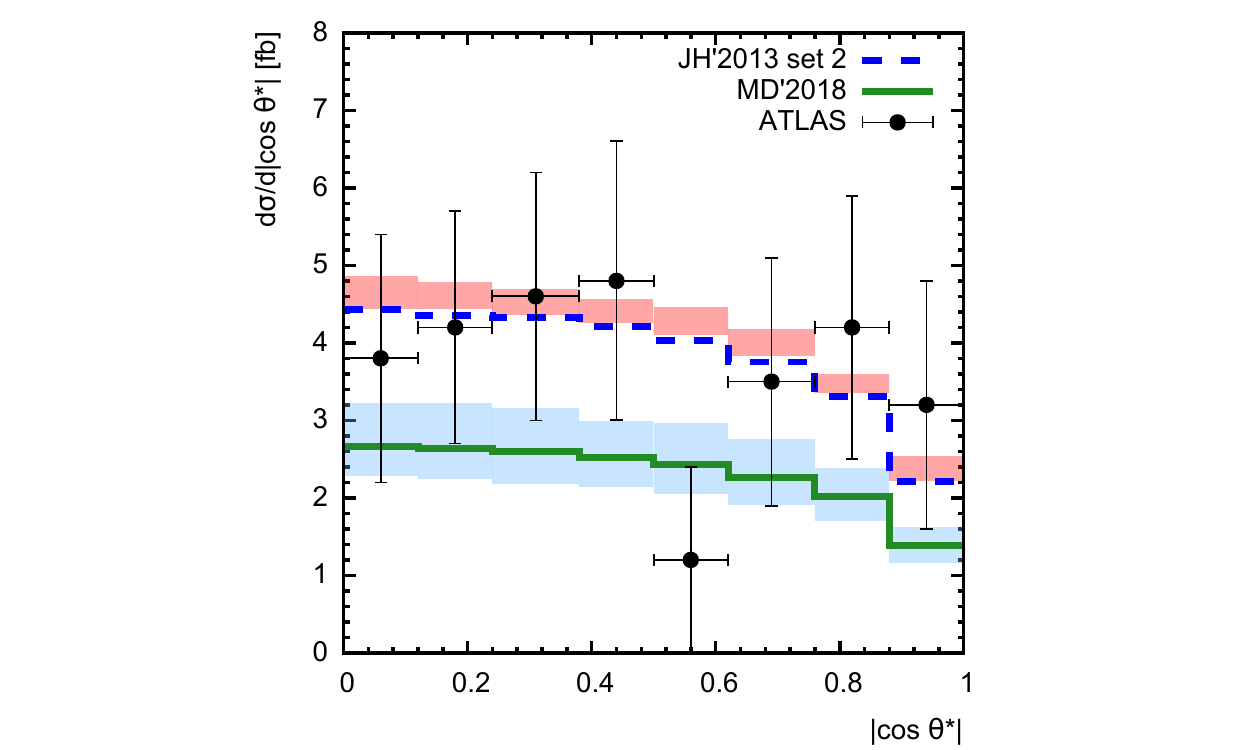}
\caption{The differential cross sections of inclusive Higgs production (in the
$H \to ZZ^* \to 4l$ decay mode) at $\sqrt s = 13$~TeV as functions of 
Higgs boson transverse momentum $p_T^H$, rapidity $|y^H|$, leading lepton 
pair decay angle $|\cos \theta^*|$ (in the Collins-Soper frame) 
and invariant mass $m_{34}$ of the subleading lepton pair.
Notation of histograms is the same as in Fig.~2. 
The experimental data are from CMS\cite{66} and ATLAS\cite{68}.}
\label{fig11}
\end{center}
\end{figure}

\end{document}